\documentclass[aps,prl,
reprint,dvipdfmx,
groupedaddress
]{revtex4-1}

\usepackage{graphicx,color}
\usepackage{amssymb}   
\usepackage{amsmath}
\usepackage{graphicx}
\usepackage{epstopdf}
\usepackage{titletoc}
\usepackage[hidelinks]{hyperref}
\usepackage{bm}
\usepackage{braket}
\usepackage{physics}
\usepackage{ulem}

\newcommand{\td}{\mathrm{d}}
\newcommand{\p}[2]{\frac{\partial {#1}}{\partial {#2}}}
\newcommand{\pa}[1]{\frac{\partial}{\partial {#1}}}
\newcommand{\de}[2]{\frac{\mathrm{d} {#1}}{\mathrm{d} {#2}}}
\newcommand{\der}[1]{\frac{\mathrm{d}}{\mathrm{d} {#1}}}

\begin{document}	
\title{Scaling Theory of Quantum Ratchet}

\author{Keita Hamamoto$^1 $}
\author{Takamori Park$^1 $}
\author{Hiroaki Ishizuka$^1 $}
\author{Naoto Nagaosa$^{1,2} $}

\affiliation{
	$^1$ Department of Applied Physics, 
	The University of Tokyo, Tokyo 113-8656, Japan \\	
	$^2$ RIKEN Center for Emergent Matter Science (CEMS), 
	Wako, Saitama 351-0198, Japan
}

\date{\today}

\begin{abstract}
The asymmetric responses of the system between the
external force of right and left directions are called ''nonreciprocal''.
There are many examples of nonreciprocal responses such
as the rectification by p-n junction.
However, the quantum mechanical wave does not
distinguish between the right and left directions as long as
the time-reversal symmetry is intact, and it is a highly
nontrivial issue how the nonreciprocal nature originates in
quantum systems.
Here we demonstrate by the quantum ratchet model, i.e.,
a quantum particle in an asymmetric periodic potential,
that the dissipation characterized by a dimensionless
coupling constant $\alpha$ plays an essential role for
nonlinear nonreciprocal response.
The temperature ($T$) dependence of the
second order nonlinear mobility $\mu_2$
is found to be $\mu_2 \sim T^{6/\alpha -4 }$ for $\alpha<1$,
and $\mu_2 \sim T^{2(\alpha -1)}$ for $\alpha>1$, respectively,
where $\alpha_c=1$ is the critical point of the
localization-delocalization transition, i.e., Schmid transition.
On the other hand, $\mu_2$ shows the behavior $\mu_2 \sim T^{-11/4}$
in the high temperature limit. Therefore, $\mu_2$ shows the nonmonotonous
temperature dependence corresponding to the classical-quantum crossover.
The generic scaling form of the velocity $v$ as a function of
the external field $F$ and temperature $T$ is also discussed.
These findings are relevant to the heavy atoms in metals,
resistive superconductors with vortices and Josephson junction system,
and will pave a way to control the nonreciprocal responses.
\end{abstract}

\maketitle

Chirality is one of the most basic subjects in whole sciences including
physics, chemistry, and biology \cite{Gardner,Tokura}. 
Most of the focus is on the symmetry of the static structures
of molecules and organs etc. However, once the motion or
flow of particles is considered, the distinction between right and left directions 
of the quantum dynamics is a highly nontrivial issue even when the
system lacks the inversion and mirror symmetries, i.e., chiral.
Classical dynamics of particle under asymmetric potential has been a deeply studied topic
	 in wide fields of science since Feynman conceived the idea of Brownian ratchet \cite{Feynman}. 
	 Researches range from molecular motor \cite{Haddou,Julicher}, colloid dynamics \cite{Rousslett}, 
	 optically trapped molecule \cite{Faucheux} to drop of mercury \cite{Gorre}.

Quantum effects on the particle dynamics under the
nonreciprocal periodic potential $V(x)$  is one of 
the most fundamental problems in condensed matter physics.
Without the dissipation, the engenstates of this problem is given 
by the Bloch wavefunctions characterized by the crystal momentum $k$
and the engenenergy $\varepsilon_n(k)$ with $n$ being the band index.
Neglecting the spin degrees of freedom, $\varepsilon_n(k)$ is symmetric 
between $k$ and $-k$, i.e., $\varepsilon_n(k) = \varepsilon_n(-k)$ as far as
$V(x)$ is real, i.e., Hermitian.
Therefore, the transport phenomena are symmetric between right and left 
directions as long as the many-body interaction is neglected \cite{Morimoto}.
This is in sharp contrast to the daily experience, which is governed by
classical mechanics, that it is more difficult 
to climb up the steeper slope compared with the gentle one. 
Especially, the role of friction is important; even at the classical
dynamics, the time-reversal symmetry and energy conservation law 
prohibit the difference between the motions to the right and left directions.
Therefore, an important question is how the dissipation brings about the
nonreciprocal transport of a quantum particle.

  Dynamics of a quantum Brownian particle in the periodic potential 
with dissipation has been
the subject of intensive studies for a long term \cite{Weiss}.
The formulation of the quantum dissipation in terms of the
coupling to harmonic bath by Caldeira-Leggett gives a way to handle
this problem in the path integral formalism \cite{Caldeira, LeggettRMP},
and the real-time formalism to calculate the influence integral 
is often used to calculated the mobility \cite{FeynmanVernon}.
Using these methods combined with the renormalization group analysis, 
the quantum phase transition is discovered at the critical value of 
the dimensionless friction $\alpha$, which separates the 
extended ground state at $\alpha< \alpha_c =1$ and 
the localized one at $\alpha> \alpha_c =1$ \cite{Schmid, GuineaPRL, 
GuineaPRB, Fisher, Zwerger, 
KanePRL, KanePRB, Furusaki}. As for the linear mobility $\mu_1$ is concerned,
it approaches to a finite value $\mu_1 \propto 1/\alpha$ when $\alpha <1$, while
$\mu_1$ vanishes as $\mu_1 \sim T^{2(\alpha-1)}$ when $\alpha>1$ in the limit 
$T \to 0$.
This transition can be regarded as that from quantum to classical
dynamics as the friction $\alpha$ increases. Therefore, it is interesting to see
how this transition affects the nonreciprocal dynamics of the quantum particle
in the asymmetric potential.

Experimentally, the quantum ratchet effects in semiconductor heterostructure with 
artificial asymmetric gating \cite{Exp1}, Josephson junction array \cite{Exp2}, and  $\varphi$ 
Josephson junction \cite{Exp3} are reported. 

Recently, the vortex flow resistance in a noncentrosymmetric superconductor is shown to 
express a large directional dichroism at the low temperature \cite{Wakatsuki}.
The classical dissipative dynamics of a point particle in the presence the asymmetric 
pinning potential is investigated as a candidate model \cite{Hoshino}, however the low 
temperature behavior is not addressed where the quantum tunneling plays a vital role.

 In this paper, we study the quantum dynamics of the particle in 
an asymmetric periodic potential with Ohmic dissipation. 
The form of the potential is for example taken as
$V(x) = V_1 \cos \left(2\pi \frac{x}{a}\right) + V_2 \sin \left(4\pi \frac{x}{a}\right)  $,
which breaks the inversion symmetry $x \to -x$.
This model describes the quantum ratchet, and several earlier works
addressed this problem \cite{Inst1,Inst2,Inst3,Inst4,Inst5,Vinokur,Peguiron,Peguiron2}. 
The instanton approach in the strong coupling limit has been employed 
in \cite{Inst2,Inst3,Inst4}, where the non-monotonous temperature
dependence of the nonlinear mobility $\mu_2$ has been obtained 
due to the crossover from temperature assisted transition to 
quantum tunneling. Here, the coherence between the tunneling events
has been neglected, which eventually becomes important in the low
temperature limit. 
Scheidl-Vinokur \cite{Vinokur} and Peguiron-Grifoni \cite{Peguiron,Peguiron2} employed the 
weak coupling perturbation theory with respect to the potential $V(x)$
and obtained the lowest order expression for the second order mobility $\mu_2 \propto V_1^2 V_2$, and the rectified velocity $v(F)+v(-F)\propto V_1^2 V_2$, respectively, in terms of the integral over the two time variables $t_1$ and $t_2$. 
However they have not carefully examined the detailed temperature dependence especially at low temperature.

\begin{figure}[t]
	\centerline{\includegraphics[width=0.35\textwidth]{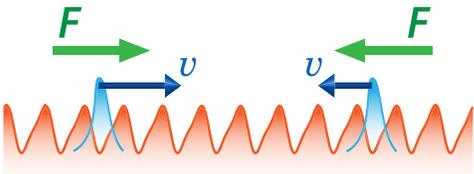}}
	\caption{ Schematic picture of the present system.
		The particle wave packet under the ratchet potential is driven by the external force  $F$ resulting in a nonreciprocal velocity; $|v(-F)|\neq |v(F)| $.
	}
	\label{Fig:Potential}
\end{figure}

Here, we rederive the general expression of the steady state velocity 
as a function of external force $F$
in the presence of the dissipation and the general form of asymmetric corrugation $V(x)$ 
in a perturbative way.
This perturbation theory is justified for $\alpha<1$, where the potential is irrelevant. We will discuss the other case $\alpha>1$ later.
The general formula for steady velocity $v(F)$ enable us to investigate the detailed 
temperature scaling for arbitrary order mobility $\mu_n$.
The dissipation is handled in terms of the 
Feynman-Vernon's influence functional technique \cite{FeynmanVernon} where 
the infinite set of harmonic oscillators with Ohmic spectral density 
$J(\omega) = \eta \omega$ are coupled bilinearly to the 
quantum mechanical point particle and integrated out. 
The lowest order perturbative expansion with respect to $V(x)$ 
allows us to compute the velocity and the mobility in the long time limit 
in the real time expression for the general strength of the dissipation, 
temperature $T$, and the external force $F$.
Since the derivation is tedious and just a straightforward generalization of earlier works \cite{Fisher,Vinokur,Eckern,Peguiron,Peguiron2}, the detail is given in the Supplemental Material (SM) \cite{SM} and we here show only the final expression. Another approach to derive the same expression is also given in SM \cite{SM}.
Throughout this paper, we set $\hbar=k_B=1$.


The zeroth order in $V$ gives $v^{(0)} = F/\eta$ and the first order correction is zero.
In the order of $V^2$, the modification to velocity is \cite{Fisher,Vinokur,Eckern,Peguiron,Peguiron2}
\begin{align}\label{eq:v2}
v^{(2)}&=  -\frac{2}{\eta}  \int_0^\infty\mathrm{d} t \sum_{k} k \left| V_{k} \right|^2   
\sin\left[\frac{F}{\eta} k t  \right] \nonumber \\
& \qquad  \times \sin \left[ \frac{1}{\pi\eta} k ^2Q_1(t) \right] 
\exp\left[  -\frac{1}{\pi\eta}   k^2Q_2(t) \right].
\end{align}
Here $V_k$ is the Fourier component of the periodic potential $V(x)$ with $k$ 
being the integer multiple of $2\pi/{a}$. $Q_1$ and $Q_2$ are \cite{LeggettRMP}
\begin{align}
Q_1(t) &= \int _0^\infty \mathrm{d} \omega \frac{J(\omega)}{\eta\omega^2 } 
\sin (\omega t) f(\omega/\gamma) \\
Q_2(t) &= \int _0^\infty \mathrm{d} \omega \frac{J(\omega)}{\eta\omega^2 } 
\left(1-\cos (\omega t) \right ) \coth\left(\frac{\omega}{2T}\right) f(\omega/\gamma). 
\end{align}
$\gamma$, being $\eta$ divided by the particle mass $M$, is the characteristic 
frequency scale in the present system. $f$ is appropriate soft cutoff function. 
Here we take $f(\omega/\gamma) =e^{-\omega/\gamma}$. This result is the same as Peguiron-Grifoni's one \cite{Peguiron,Peguiron2} and reduces to 
the Scheidl-Vinokur's result \cite{Vinokur} in the small $F$ limit and to 
Fisher-Zwerger's result \cite{Fisher} if we take only $k=\pm\frac{2\pi}{a}$. Note here 
that as the effect of the asymmetry of the potential $V(x)$ is missing in this formula, 
this result in nothing to do with the ratchet effect therefore $v^{(2)}$ is the odd 
function of $F$.
To clarify the low temperature behavior of $v^{(2)}$, the asymptotic forms of $Q_1$ 
and $Q_2$ for $t ,T^{-1} \gg \gamma^{-1}$ are important;
\begin{align}
Q_1(t) &=  \tan ^{-1} (\gamma t)\rightarrow const.\\
Q_2(t) &= \log\left( \left[1+(\gamma t)^2\right]^{1/2}\left|
\frac{\Gamma(1+
\frac{T}{\gamma})}{\Gamma(1+\frac{T}{\gamma}+iTt)}\right|^2 \right)  \nonumber \\
&\rightarrow \log (\gamma t)  + \log \left( \frac{\sinh(\pi T t)}{\pi T t}\right)
\label{eq:asymptoticsQ2}
\end{align}
with $\Gamma(\cdot)$ being the Gamma function.
From these asymptotic behaviors, when expanded in $F$, the $n$-th order 
term of $v^{(2)}$ scales in the leading order as
\begin{align}
v^{(2)} \sim T^{\frac{2}{\alpha}-1-n}  F^n 
\end{align}
in the order of $F^n$ with $n$ being odd integers. Here widely used 
dimensionless dissipation strength is 
\begin{align}
\alpha=\frac{\eta a^2}{2\pi}.
\end{align}

In the third order of $V$'s, where the quantum ratchet 
effect appears, we similarly have \cite{Fisher,Vinokur,Eckern,Peguiron,Peguiron2}
\begin{widetext}
\begin{align}
v^{(3)}&= \frac{4}{\eta}  \int_0^{\infty} \mathrm{d} t_1  \int_{0}^{\infty} 
\mathrm{d} t_2 \sum_{\substack{ k_1,k_2,k_3 \\ k_1+k_2+k_3=0}} k_1  \nonumber \\
&\times \left( \Re\left[  V_{k_1}V_{k_2} V_{k_3} \right] \sin\left[ \frac{F}{\eta} 
\left(  k_1 t_1 - k_3 t_2\right) \right]+\Im\left[  V_{k_1}V_{k_2} V_{k_3} \right]  
\left( \cos\left[ \frac{F}{\eta} \left(  k_1 t_1 - k_3 t_2\right) \right] -1\right) \right)
\nonumber\\
&\times    \exp\left[ \frac{1}{\pi\eta} \left(  k_1 k_2 Q_2(t_1) + k_2 k_3 Q_2(t_2) 
+k_3k_1Q_2(t_1+t_2)  \right) \right] \sin \left[ \frac{1}{\pi\eta}  k_1 k_2 Q_1(t_1)\right]  
\sin \left[ \frac{1}{\pi\eta} \left(  k_2k_3 Q_1(t_2) + k_3 k_1Q_1(t_1+t_2) \right)  \right] .
\label{eq:v3}
\end{align}
\end{widetext}

This result reduces to 
the Scheidl-Vinokur's result \cite{Vinokur} in the order of $F^2$ and reproduces the Peguiron-Grifoni's result for the rectified velocity $v(F)+v(-F)$ in the presence of up to the second harmonic potential; $k=\pm\frac{2\pi}{a},\pm\frac{4\pi}{a}$ \cite{Peguiron,Peguiron2}.
Although the expression is rather complex, we can see the behavior in the low 
temperature limit by the power-counting of the integrand using the asymptotic 
forms as follows. We see from Eq.(\ref{eq:asymptoticsQ2}) that the exponential 
of $ -Q_2(t)$ function gives us a power of $t$ and the large $t$ 
cutoff of the form $\exp[-\pi Tt]$ at finite temperature. Thus we are allowed 
to count the power at zero temperature and cutoff the integral domain 
$[0,T^{-1}]$ to see the $T$ dependence at low temperature.

\begin{figure}[b!]
	\centerline{\includegraphics[width=0.25\textwidth]{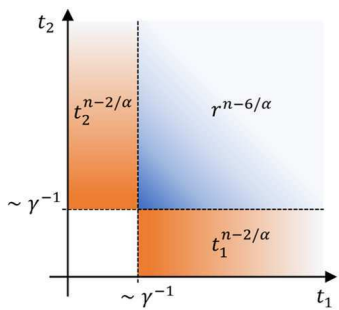}}
	\caption{
		Asymptotic behavior of the integrand of $n$-th order expansion with respect to 
		$F$ of Eq.(\ref{eq:v3}) in each region in the $t_1$-$t_2$ plane. 
		As the leading order contributions from the orange regions cancel out among the terms, the blue region determines the temperature behaviors.
	}
	\label{Fig:Integrand}
\end{figure}

The dominant contribution to the integral originates from 
$(k_1,k_2,k_3) = \pm \frac{2\pi}{a}(1,1,-2)$ and its permutations. By means of the 
polar coordinate $(r,\theta)$, the integral is 
$F^n \int r \mathrm{d} r \ r^{n-\frac{6}{\alpha}} \sim T^{\frac{6}{\alpha}-2-n} F^n$. 
On the other hand, if we fix one of the variables, say $t_1$, the integral behaves as 
$F^n \int\mathrm{d} t_2 \ t_2^{n-\frac{2}{\alpha}} \sim T^{\frac{2}{\alpha}-1-n} F^n$. 
	Although the latter contribution seems to dominate the former one at low temperature for $\alpha<4$, the closer inspection shows that the summation over $k_1,k_2,k_3$ causes an exact cancellation of these leading order contributions. The proof of this cancellation is given in SM \cite{SM} and numerical calculations support this cancellation up to $12$ digits in double precision calculations. Thus, the low temperature exponent is governed by the sub leading contributions;
\begin{align}\label{v3scaling}
v^{(3)} \sim T^{\frac{6}{\alpha}-2-n}  F^n
\end{align}
in the order of $F^n$ with $n$ being a positive integer.

\begin{figure}[t]
	\centerline{\includegraphics[width=0.5\textwidth]{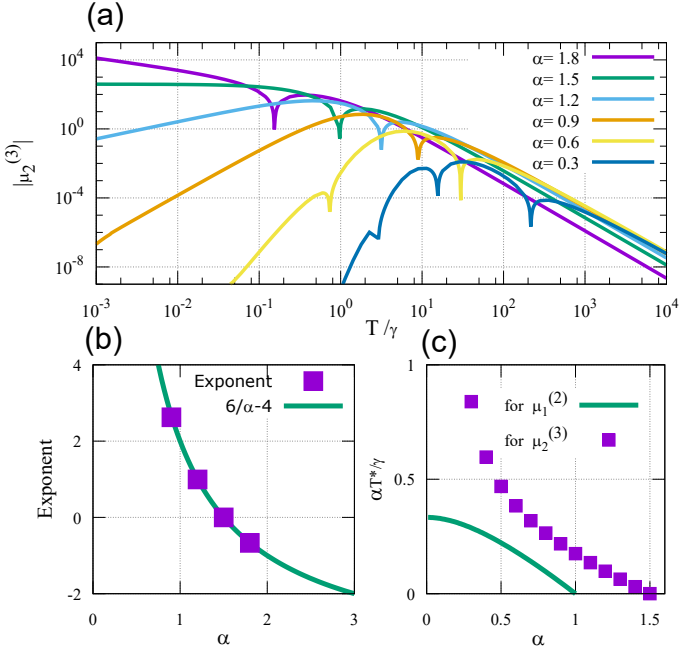}}
\caption{ Temperature dependence of the second order mobility $\mu_2^{(3)}$.
		(a) The second order mobility $\mu_2^{(3)}$ is evaluated from Eq.(\ref{eq:v3}) for the asymmetric potential $V(x)  = V_1\cos(2\pi x/a) +V_2 \sin(4\pi x/a)$ with $V_2=V_1/4$ for each value of $\alpha$.
		There are two power law regions with different exponents; $6/\alpha-4$ for low temperature region and $-11/4$ for high temperature region. The quantum-to-classical crossover region with sigh changes (cusps) in between is also seen. 
		For $\alpha>1$, as the pertubative expansion with respect to the potential fails 
		and the system goes to the localized phase, $\mu_2^{(3)}$ vanishes at zero temperature. 
		Therefore there must be another crossover point $T^{\ast\ast}$ at low temperature 
		where the perturbative treatment breaks down.
		(b) The low temperature power low exponent of $\mu_2^{(3)}$ which clearly follows 
		the asymptotic form $\mu_2^{(3)}\propto T^{6/\alpha-4}$.
		(c) The crossover temperature $T^\ast$ defined by peak positions in (a).
		The green line is that for $\mu_1^{(2)}$ evaluated from Eq.(\ref{eq:v2}).
	}
	\label{Fig:Mu2}
\end{figure}

The numerical evaluation of second order mobility $\mu_2^{(3)}$ which is given by the
expansion of Eq.(\ref{eq:v3}) with respect to $F$ 
depicted in Figs.\ref{Fig:Mu2}(a) and (b) clearly show
temperature dependence as described by Eq.(\ref{v3scaling}) at low temperature. 
For $0< \alpha<3/2 $, $\mu_2$ turn to decrease as decreasing temperature around $T=T^\ast \sim \gamma$. 
This is a peculiar behavior of the present system which can be captured in real experiments.
For $\alpha>1$, the potential is a relevant operator, and therefore the pertubative expansion with respect to the potential 
diverges towards the low temperature.
In this case, the system is in the localized phase, and therefore $\mu_2^{(3)}$ must vanish at the zero temperature. 
This indicates the existence of another crossover temperature $T^{\ast\ast}$, which can be lower than $T^\ast$ when the potential is weak enough.
In the view point of renormalization group (RG) analysis, the potential $V$ scales as $V(\Lambda) =V(\Lambda_0) (\Lambda/\Lambda_0)^{1/\alpha-1}$ for the high energy cutoff $\Lambda$ \cite{Fisher}. The cutoff is truncated at $\Lambda\sim T$ at finite temperature therefore we can estimate the crossover temperature as $V(\Lambda_0) (T^{\ast\ast} /\Lambda_0)^{1/\alpha-1}\sim T^{\ast\ast}$.

The higher crossover temperature deduced from the peaks of Fig.\ref{Fig:Mu2}(a) is shown in Fig.\ref{Fig:Mu2}(c)
together with that for the linear mobility $\mu_1^{(2)}$ evaluated from Eq.(\ref{eq:v2}).
The crossover temperature for $\mu_2^{(3)}$ is always larger than that for $\mu_1^{(2)}$ but is comparable.
Thus we can conclude that the crossover observed in $\mu_2^{(3)}$ is the quantum to classical crossover
as known for $\mu_1^{(2)}$. Note that the peaks in $\mu_2^{(3)}$ for small $\alpha$ is not clear due to many sign changes in the crossover region.
		
This low temperature dependence is in contrast to the saturating behavior discussed in ref.\cite{Vinokur} where a nontrivial approximation is made in the evaluation of $Q_2$, which fails to capture the quantitative behavior of $\mu_2^{(3)}$.	
For the higher temperature, $\mu_2^{(3)}$ decreases equally irrespective of 
$\alpha$ as $\mu_2^{(3)}\sim T^{-11/4}$ whose derivation is given in SM \cite{SM}.
This value is slightly different from $T^{-17/6}$ obtained in ref.\cite{Vinokur}. This discrepancy is due to the difference of the choice of cutoff function $f(\omega/\gamma)$ as discussed in SM.
In the intermediate temperature, the crossover-like behavior and some sign changes are observed as
pointed out by ref.\cite{Vinokur}.


For $\alpha <1$, the perturbative treatment of the potential $V$'s is appropriate. And the leading order terms leads to the scaling form in the low temperature limit as

\begin{align}
v& = \frac{F}{\eta} - F^{2/\alpha -1} f_o^<(F/T)  - F^{6/\alpha -2} f_e^<(F/T)  \notag \\ 
&= \frac{F}{\eta} - T^{2/\alpha -1} g_o^<(F/T) - T^{6/\alpha -2} g_e^<(F/T)
\label{eq:scaling1}
\end{align} 
where $f_o^<,g_o^<$ are odd functions while $f_e^<,g_e^<$ are even.
The basis of this scaling is that the velocity vanished in the limit $F \to 0$, 
which is given by the integral region of large 
time variable $t  \gtrsim 1/F$. Note that only the asymptotic behavior of the 
integrand at large time variable determines the scaling behavior for the velocity $v$ 
itself, while the expression for the coefficient of $F_n$ for the velocity $v$
does not appear so. Therefore, the divergence of the
nonlinear mobility as $T \to 0$ does not mean the
divergence of $v$, but the functional form becomes non-analytic 
at the zero temperature $T =0$. 
In Eq.(\ref{eq:scaling1}), the functions $g_o^<,g_e^<$ are an analytic functions of their argument
$F/T$ since the perturbative expansion is always possible when $F \ll T$,
while $f_o^<,f_e^<$ are not. Trivially, they are related by 
$f_i^{<}( \eta) = \eta^{ 1 - 2/\alpha} g_i^<(\eta)$ with $i\in \{e,o\}$.
The role of the nonreciprocal potential, i.e., $V_2$, is to introduce 
the even component $g_e^<$.
One can easily see that the second order nonlinear mobility
$\mu_2$ scales as $\mu_2 \sim T^{6/\alpha -4}$. 
Furthermore, the generic odd (even) nonlinear mobility
of $n$-th order scales as 
$\mu_n \sim T^{2/\alpha -n -1 }$ ($\mu_n \sim T^{6/\alpha -2 -n }$) for $\alpha<1$,
and it diverges when $2/(n+1) < \alpha < 1$ ($6/(n+2) < \alpha < 1$) while
it vanished otherwise in the limit $T \to 0$.
Note here that the $I$-$V$ relation of the Tomonaga-Luttinger liquid (TLL) system under weak asymmetric potential $I\sim V^{6g-2}$ with $g$ being the Tomonaga-Luttinger's interaction parameter, is shown in ref. \cite{Feldman} which is analogous to the $f_e^<$ term in eq.(\ref{eq:scaling1}).
There are many well-known similarities between the present system and the TLL system \cite{KanePRB,KanePRL} and some of them are exemplified in SM \cite{SM}.

From the viewpoint of the RG, 
$V_1$ is irrelevant for $\alpha <1$ while becomes relevant for
$\alpha>1$. Similarly $V_2$ is irrelevant for $\alpha <4$, and 
becomes relevant for $\alpha>4$. Naively, this might lead to 
the critical $\alpha$ being $4$ for the nonreciprocal mobility.
However, the RG procedure generates the composite operator 
$V_1 V_2$, which includes $\sin \left(2\pi \frac{x}{a}\right) $, 
which has the same scaling dimension as $V_1$.
This fact is reflected in each term of the double time integral where
the dominant contribution comes from the region where
one of $t_1$ and $t_2$ is finite, and the asymptotic 
behavior is basically given by the one-dimensional integral over time.
However, the combination of $\cos(2\pi x/a)$ and $\sin(2\pi x/a)$ simply shifts the potential leaving the inversion symmetry intact. This is the reason why the cancellation occurs for the leading order terms 
$\propto T^{ 2/\alpha-1-n}$ in $v^{(3)} $

Now we turn to the case of $\alpha>1$, where $V$'s are relevant and scale 
to larger values \cite{Schmid}. In this case, the tunneling $t$ between the 
potential barrier is the irrelevant operator, and the perturbation theory 
in $t$ should be employed \cite{KanePRL,KanePRB}.
The question is how the asymmetry of the potential 
enters the problem. For this purpose, let us consider the
tilting of the potential under the external field $F$.
Due to the asymmetry of the potential, 
the change in the potential barrier linear in $F$ 
exists, which results in the $F$-dependence of 
$t$, i.e., $t(F) = t + \gamma F$.
This $t(F)$ is used for the calculation of $v$ 
in the lowest perturbation, which results in 
\begin{align}
v& = t(F)^2 F^{2 \alpha -1} f_o^>(F/T) \notag \\ 
&=  t(F)^2 T^{2 \alpha -1} g_o^>(F/T),
\label{eq:scaling2}
\end{align} 
where $g_o^>(F/T)$ is the odd function of its argument, i.e., it contains
only the odd order term in the Tailor expansion.
Therefore, the second order nonlinear mobility $\mu_2$ scales with 
$ T^{2 (\alpha - 1)}$ similarly to the linear mobility $\mu_1$, and goes to 0 as
$T \to 0$.

For the check of the scaling form Eq.(\ref{eq:scaling2}) also in the strong 
coupling regime where potential terms are relevant operators,
we calculated a temperature dependence of the linear and the 
third order mobility in the perturbation in $t$.
As shown in detail in SM \cite{SM}, by the perturbation with respect 
to the tunneling amplitude, they precisely follow the expected power 
law as Eq.(\ref{eq:scaling2}).

Lastly, we comment on the array of resistively shunted josephson juntion model,
which is a direct generalization of the present system to higher dimensions.
This model, composed of the superconducting islands 
connected by Josephson couplings with symmetric cosine potential and the 
shunting Ohmic dissipation, 
is a promising candidate to explain the low temperature behavior 
of the thin film of granular superconductors \cite{Chakravarty,Kapitulnik}.
It is shown that the model shows a quantum phase transition between 
coherent (superconducting) and disordered (normal) states at 
$\alpha = h/(4e^2 R) =1/z_0$ 
where $R$ is the shunting resistance and $z_0$ is the half of the 
coordination number of the lattice of islands \cite{Chakravarty}.
If we introduce a asymmetric potential to the Josephson phase, 
the nonlinear transport coefficients of the system should follow the 
present scaling form. 
One difference is that the current in the Josephson array acts as a tilting 
to the potential
while the resulting time derivative of the Josephson phase is the voltage drop, 
 therefore nonlinear resistivity, instead of mobility, follows the scaling given in the present paper. 
 Another difference is the absence of the voltage drop for $z_0 \alpha >1$ 
due to the superconductivity. 
Thus we can conclude that $n$-th order resistivity with odd (even) $n$ scales as 
$R_n\sim T^{2/(z_0\alpha)-n-1}$ ($R_n\sim T^{6/(z_0\alpha)-n-2}$) 
and diverges when $2/(n+1) < z_0\alpha < 1$ ($6/(n+2) < z_0\alpha < 1$) at zero temperature.

 In summary, we have studied the role of dissipation in the 
nonreciprocal transport of quantum particle in the asymmetric 
periodic potential, i.e., quantum Ratchet model.
We have derived the general expression of the steady state velocity $v$ for the general value of the dissipation $\alpha$, force $F$, temperature $T$, and shape of the periodic potential $V(x)$
 and found different scalings behavior at low temperature depending on the even and odd powers of $F$.
This results can be applied to various
situations such as asymmetric Josephson junction array, motion 
of heavy atoms in noncentrosymmetric crystal, and 
vortex motion in noncentrosymmetric superconductors.

{\bf Acknowledgment.} ---
We are grateful to A. J. Leggett for fruitful discussions.
N.N. was supported by Ministry
of Education, Culture, Sports, Science, and Technology
Nos. JP24224009 and JP26103006, the Impulsing Paradigm
Change through Disruptive Technologies Program of Council
for Science, Technology and Innovation (Cabinet Office,
Government of Japan), and JST CREST Grant Number JPMJCR1874, and  JPMJCR16F1, Japan.
K.H. was supported by JSPS through a research fellowship for young
scientists and the Program for Leading Graduate Schools (MERIT)

\clearpage
\onecolumngrid

\appendix

\vspace{15pt}
\begin{center}
	{\large \bf ---Supplementary Materials---}
\end{center}

\setcounter{figure}{0}
\setcounter{equation}{0}
\setcounter{table}{0}
\renewcommand{\thefigure}{S\arabic{figure}}
\renewcommand{\theequation}{S\arabic{equation}}
\renewcommand{\thetable}{S\Roman{table}}
\baselineskip=6mm


\section{S1. Similarity to the Tomonaga-Lutinger liquid system}
There are many similarities between the quantum Brownian particle studied in the main text and the Tomonaga-Luttinger liquid (TLL) \cite{KanePRB,KanePRL}.
The former system is characterized by the dissipation strength $\alpha$ while
the interaction parameter $g$ determines behaviors of the latter system.
The effective action of each system is equivalent to the other with the
correspondence $g = 1/\alpha$ .
Combined with the self-duality of this problem \cite{Schmid,Fisher,Peguiron},
where the potential problem with parameter $\alpha$ is mapped onto
the tunneling one with $1/\alpha$, one can also find a correspondence
between the potential problem of quantum Brownian motion
and the weak tunneling problem in TLL.
In this section, we will outline the similarities of the two models. 

Since the problem of barriers in TLL have been extensively studied already, here
we will curtail detailed derivations and only discuss the similarities between the two models. The Euclidean action of a TLL is,
\begin{equation}
S=\int\dd x\dd\tau\frac{vg}{2}\qty((\nabla\phi)^2+\frac{1}{v^2}(\partial_\tau\phi)^2),
\end{equation}
where $g$ is the interaction parameter of the TLL. For $0<g<1$, the interaction
is repulsive and for $g>1$, the interaction is attractive.
The effective action of the TLL with a strong barrier, or equivalently a small tunneling, can be obtained by considering
two semi-infinite TLL that are connected by a perturbative hopping term with
strength $t_n$ at $x=0$. After integrating out the field for $x\neq0$, the effective
Euclidean action is found to be,
\begin{equation}
S[\phi]=g\sum_{\omega_n}\abs{\omega_n}\abs{\phi(\omega_n)}^2
+\frac{1}{2}\sum_{n=-\infty}^\infty t_n\int_0^\beta\dd\tau e^{i2n\sqrt{\pi}\phi(\tau)},
\end{equation}
where  $\phi(\tau)=[\phi_R(\tau,x=0)-\phi_L(\tau,x=0)]/2$ is the difference between the bosonic
phase field of the left and right semi-infinite TLL at $x=0$ and $t_n$ is the
hopping strength of $\abs{n}$ electrons being hopped to the right when $n>0$ or
to the left when $n<0$.

A renormalization group analysis of the action shows that the hopping $t_n$
is irrelevant for the interaction parameters $g<1$ which corresponds to repulsive
interaction, so in this discussion, we limit ourselves to the case of repulsive
interaction where we can safely employ perturbative methods.
The current is obtained by first inserting a `vector potential' $a(t)$---such that
the applied voltage is $V(t)=\partial_ta(t)$---into the argument of the hopping
term, taking the functional derivative of the partition function, and then
deforming the contour integral from the negative imaginary axis to the real axis.
In imaginary time, the contribution to the current from the term third order in
hopping strength is \cite{KanePRB},
\begin{equation}
I^{(3)}(\tau)=-\frac{i}{16}\sum_{n_1+n_2+n_3=0}n_3t_{n_1}t_{n_2}t_{n_3}
\int_0^\beta\dd\tau_1\int_0^\beta\dd\tau_2
P_{-2n_1n_2/g}(\tau_1-\tau_2)P_{-2n_1n_3/g}(\tau_1-\tau)P_{-2n_2n_3/g}(\tau_2-\tau),
\end{equation}
where
\begin{equation}
P_{\lambda}(\tau)=\qty(\frac{\pi\tau_c/\beta}{\sin(\pi\tau/\beta)})^{\lambda},
\end{equation}
and $\tau_c$ is a short-time cutoff.
The analytic continuation is performed by using the closed time path contour which
extends from $t_i=-\infty\rightarrow t_i=t$ then $t_i=t\rightarrow t_i=-\infty$
while being careful with how the imaginary time ordering now becomes a contour
ordering. The third order contribution in hopping strength to the current in real time is,
\begin{align}
I^{(3)}=
\frac{1}{2}\int_0^\infty\dd t_1\int_0^\infty\dd t_2
\sum_{\substack{n_1+n_2+n_3=0\\n_i\in\mathbb{Z}}}
&n_1t_{n_1}t_{n_2}t_{n_3}
\exp[\frac{2}{g}\qty(n_1n_2Q(t_1)+n_2n_3Q(t_2)+n_1n_3Q(t_1+t_2))]\notag\\
&\times\sin\qty[\frac{\pi}{g}n_1n_2]\sin\qty[\frac{\pi}{g}(n_2n_3+n_1n_3)]
\sin\qty[V(n_1t_1-n_3t_2)],
\label{eq:TLL_third_order}
\end{align}
where,
\begin{equation}
Q(t)=\log\left(\frac{t}{\tau_c} \right) + \log\qty(\frac{\sinh(\pi Tt)}{\pi Tt}).
\end{equation}
Notice the similarities with the expression for the third order contribution in $V$ to the velocity
of the particle in the ratchet potential. Eq. \eqref{eq:TLL_third_order} can be
obtained from that equation by substituting $Q_1(t),Q_2(t)$ with its asymptotic form at
long time and low temperature, and
only looking at terms that are odd in $F$. 
There is a clear correspondence
between the interaction parameter of TLL, $g$, and the dissipation strength
of the quantum Brownian particle, $\alpha$.
As the weakly linked TLL with parameter $g$ is mapped to the TLL under weak potential with $1/g$, the relation between parameters is $g=1/\alpha$ as expected.

Note that in ref. \cite{Feldman}, the authors showed $I\sim V^{6g-2}$ in the weak potential TLL system. This is corresponding to one of our central results $v\sim F^{6/\alpha-2}$ at the zero temperature which is governed by the $f_e^<$ term in eq.(\ref{eq:scaling1}) in the main text.

\section{S2. Conductance of a weak link in Tonomaga-Luttinger liquid}
In this section, we numerically compute the first and third order conductance
of a weak link in an interacting TLL. We find that the current
is consistent with the scaling form,
\begin{equation}
I=T^{2/g-1}g\left(\frac{V}{T}\right)\,.
\label{eq:llScalingForm}
\end{equation}

We model a weak link or a high barrier in an interacting TLL by
adding a hopping term of strength $t$ between two disconnected semi-infinite
TLL with interaction strength $g$. The effective long-range action
of this model is given by \cite{KanePRB},
\begin{equation}
S[\phi]=g\sum_{i\omega_n}\abs{\omega_n}\abs{\phi(\omega_n)}^2+t\int\dd\tau\cos[2\sqrt{\pi}\phi(\tau)]\,.
\label{eq:llEffectiveAction}
\end{equation}
Kane et. al. derived an expression up to second order in $t$ for the current across
the weak link when a voltage $V$ is applied \cite{KanePRL},
\begin{equation}
I\propto t^2 (1-e^{-\beta V})\tilde{P}(V)\,,
\label{eq:llCurrent}
\end{equation}
where, $\tilde{P}(V)$ is the Fourier transform of $P(t)$ defined by,
\begin{equation}
\ln P(t) = -\int_0 ^{\omega_c}\dd\omega \frac{2}{\omega g}
\qty[ \coth \qty( \frac{\omega}{2T} )(1 - \cos(\omega t)) + i\sin ( \omega t) ]\,.
\end{equation}
By introducting an exponential cutoff function $e^{-\omega/\omega_c}$ and extending the upper integration
limit to infinity, $P(t)$ can be neatly expressed as,
\begin{equation}
P(t)=\exp \left[-\frac{2}{g} \left(i Q_1(t)+Q_2(t)\right) \right]\,,
\end{equation}
where,
\begin{align}
Q_1(t)&=\tan^{-1}(\omega_c t)\,,\\
Q_2(t)&=\frac{1}{2}\ln(1+\omega_c^2t^2)+\ln(\frac{\sinh(\pi T t)}{\pi Tt})\,.
\end{align}
The Fourier transform, $\tilde{P}(V)$ is numerically computed using a slightly
modified version of the algorithm outlined by Thakkar et. al. \cite{Thakkar}.

The first and third order conductances may be computed by approximating the
current, as a function of $V$, as a sum of Chebyshev polynomials and then
taking the first and third order coefficients of the full polynomial expansion
\cite{Recipes}. In this way, the first and third order conductances were
numerically calculated for different values of $T$ and $g<1$.

\begin{figure}
	\centerline{\includegraphics[width=\textwidth]{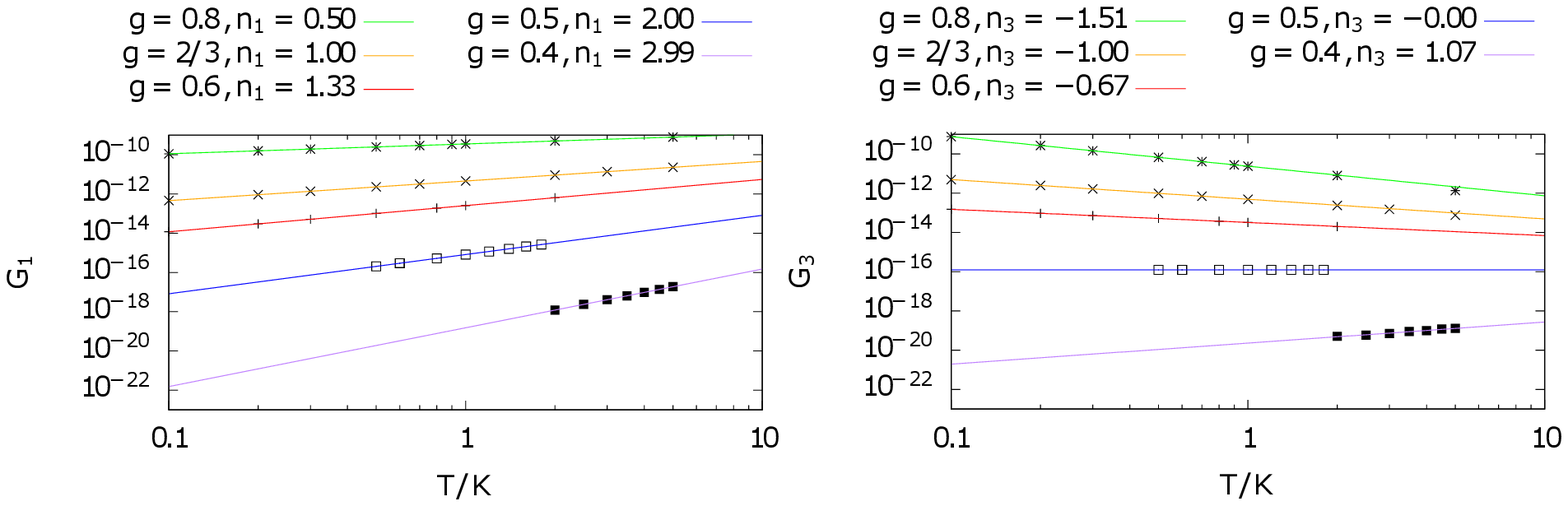}}
	\caption{Log-log plot of the temperature dependence of the first ($G_1$)
		and third order ($G_3$) conductances for $g=0.4\,,0.5\,,0.6\,,2/3\,,$ and $0.8$.
		The data points were numerically computed and the lines are regression lines.
		The linearity of the data suggests a power law temperature dependence. $n_1\,,
		n_3$ give the slopes of the lines and the exponents of the
		temperature dependence.
	}
	\label{fig:llLogLog}
\end{figure}

Figure \ref{fig:llLogLog}  is a log-log plot of the computed temperature
dependence of the first and third order conductances for different values of $g$.
The linear relationship indicates a temperature dependence that conforms to a 
power law of the form,
\begin{align}
G_1&\propto T^{2/g-2}\,,\\
G_3&\propto \frac{G_1}{T^2}\propto T^{2/g-4}\,.
\end{align}
This is consistent with \ref{eq:llScalingForm},
\begin{align}
I&=G_1 V +G_3 V^3+\mathcal{O}(V^5)\notag\\
&=a T^{2/g-2}V+bT^{2/g-4}V^3+\mathcal{O}(V^5)\notag\\
&= T^{2/g-1}g\qty(\frac{V}{T})\,,
\label{eq:llScalingForm2}
\end{align}
where $a,b\in\mathbb{R}$ are arbitrary constants and $g(\cdot)$ is an odd 
analytic function.

\section{S3. Derivation of equation (1) and (8)}
In this section, we show the derivation of a general formula of the steady velocity in a tilted periodic potential with ohmic dissipation.
This is a direct generalization of the formula developed by Fisher and Zwerger \cite{Fisher}, in which only the symmetric sinusoidal potential $V(x)=V_1\cos\left(\frac{2\pi}{a}x\right)$ is considered, to the generic periodic potential. Similar generalization is done by Peguiron and Grifoni \cite{Peguiron,Peguiron2} where the rectified velocity $v(F)+v(-F)$ for the $V(x)=V_1\cos\left(\frac{2\pi}{a}x\right)+V_2\sin\left(\frac{4\pi}{a}x\right)$ is considered.

\subsection{S3-1. Influence functional formalism}
In this section, we briefly review the influence functional formalism just by following the Fisher and Zwerger. For the detail, see it and references therein.

In Feynmann-Vernon's influence functional theory, the density matrix of system is obtained by taking a partial trace, by degrees of freedom of harmonic bath, of that of the total one;
\begin{equation}
\rho(t)=\mathrm{Tr}_{\mathrm{bath}} \left[\rho_{\mathrm{tot}}(t)\right].
\end{equation}
In the coordinate representation,
\begin{equation}
\Braket{q|\rho(t)|q'} = \int \td q_0 \int\td q_0'\Braket{q_0 | \rho(0) | q_0'} J (q,q',t|q_0,q_0',0)
\end{equation}
with $J$ being given by the double path integral;
\begin{equation}
J (q,q',t|q_0,q_0',0) = \int_{q_0}^{q} \mathcal{D} q \int_{q'_0}^{q'}\mathcal{D}q' \exp\left[ i \left[ S(q)-S(q')\right] + i \Phi (q,q') \right].
\end{equation}
The action is
\begin{equation}
S(q)=\int_0^t \td t' \left[ \frac{M}{2}\dot{q}^2 - U(q) \right] 
\end{equation}
with tilted periodic potential;
\begin{equation}
U(q)=V(q)-Fq = \sum_{k} V_k e^{i k q} -Fq.
\end{equation}
Momentum $k$ is an integer multiple of $\frac{2\pi}{a}$. The influence phase $\Phi$ is
\begin{align}
i\Phi(x,y) = &-i\int_0^t \td t' \int_{t'}^t \td s \ 2 x(t')\alpha_I(s-t') y(s) -i M (\Delta \omega)^2 \int_0^t \td t'  x(t')y(t') \nonumber\\
&- \int_0^t \td t' \int_0^{t'} \td s \ y(t')\alpha_R(t'-s) y(s) \label{Phi}
\end{align}
with
\begin{equation}
x=(q+q')/2 , \qquad y=q-q'.
\end{equation}
The integral kernels are
\begin{align}
\alpha_I(t) &=-\int_0^\infty \frac{\td \omega}{\pi } J(\omega) \sin\omega t \\
\alpha_R(t)&=\int_0^\infty \frac{\td \omega}{\pi } J(\omega) \cos\omega t \coth\left(\frac{\hbar \omega}{2T}\right)  \\
\frac{1}{2}M (\Delta \omega)^2 &= \int_0^\infty \frac{\td \omega}{\pi } \frac{J(\omega)}{\omega}
\end{align}
The specialized expression for the Ohmic dissipation $J(\omega)=\eta \omega$ is
\begin{equation}
i\Phi(x,y) = i \eta \int_0^t \td t' x(t')\dot{y}(t') -i\eta x(t) y(t) -S_2[y]
\end{equation}
with
\begin{equation}
S_2[y] = \int_0^t \td t' \int_0^{t'} \td s \ y(t') \alpha_R(t'-s) y(s) .
\end{equation}

The potential term in action is nonlinear therefore we expand as
\begin{align}
\exp\left[ i \left[ S(q)-S(q')\right]\right] &= \exp\left[ i\int_0^t \td t'\ \frac{M}{2} \left( \dot{q}^2 -\dot{q}'^2\right)  -\sum_{k} V_k e^{i k q} +Fq+\sum_{k} V_k e^{i k q'} -Fq'\right] \nonumber \\
&= \exp\left[i \int_0^t \td t'\  M \dot{x} \dot{y}  +Fy\right]\nonumber \\
& \qquad  \times \left[ \sum_{n=0}^\infty  (-i)^n \int_0^t \td t_1\cdots\int_0^{t_{n-1}} \td t_n \sum_{k_1k_2\cdots k_n} \left(\prod_{i=1}^n V_{k_i} \right) \exp\left[ -i \int_0^t \td t'{\rho(t') q(t')} \right] \right] \nonumber \\
& \qquad  \times\left[ \sum_{m=0}^\infty i^m \int_0^t \td t'_1 \cdots\int_0^{t'_{m-1}} \td t'_m \sum_{k'_1k'_2\cdots k'_m} \left( \prod_{i=1}^m V_{k'_i} \right)  \exp\left[ +i \int_0^t \td t'{\rho'(t') q(t')} \right] \right] 
\end{align}
where we have defined two "charge" densities
\begin{equation}
\rho(t') = -\sum_{i=1}^n k_i \delta (t'-t_i) , \qquad \rho'(t') = \sum_{i=1}^m k'_i \delta (t'-t'_i).
\end{equation} 
Thus the probablity density is written as
\begin{align}
P(x,t) &= \Braket{x|\rho(t)|x}\nonumber \\
&=\sum_{n,m=0}^\infty  (-i)^n i^m \int_0^t \td t_1\cdots\int_0^{t_{n-1}} \td t_n  \int_0^t \td t'_1 \cdots\int_0^{t'_{m-1}} \td t'_m \sum_{k_1\cdots k_n,k_1'\cdots k'_n} \left(\prod_{i=1}^n V_{k_i} \right)  \left( \prod_{i=1}^m V_{k'_i} \right)  \nonumber \\
& \times \int \td x_0 \int\td y_0'\Braket{x_0+\frac{y_0}{2} | \rho(0) | x_0-\frac{y_0}{2}} G
\end{align}
with $G$ being
\begin{align}
G&=\int_{x_0}^x \mathcal{D}x\int_{y_0}^0  \mathcal{D}y
\exp\left[ i\int_0^t \td t'\  M \dot{x} \dot{y}  +Fy  -(\rho-\rho') x -\frac{1}{2}(\rho + \rho') y + \eta x \dot{y} \right] \nonumber \\
& \qquad \qquad \exp\left[-i\eta x(t) y(t) -S_2[y] \right] \nonumber \\
&= A(t) \exp\left[ i\int_0^t \td t'\  \left[ F  -\frac{1}{2}(\rho + \rho')  \right] y \right]  \exp\left[-M\left. x\dot{y}\right|_0^t -S_2[y] \right] 
\end{align}
Now, coordinate $y$ is restricted to the solution of $ M \ddot{y} - \eta \dot{y} = \rho'-\rho. $ We have used $y(t'=t)=0$ due to the boundary condition. The prefactor is $A(t) = \frac{M}{2\pi d(t)}$ with $d(t') = \gamma^{-1}(1-e^{-\gamma t'})$.

By the same argument as in Fisher-Zwerger \cite{Fisher}, we see in the $t \rightarrow \infty$ limit, the finite contribution comes from the configuration with
\begin{equation}
\int_0^t \td t' (\rho-\rho') = 0 \qquad \Leftrightarrow \qquad \sum_{i=1}^n k_i + \sum_{i=1}^m k'_i = 0
\end{equation}
which is the "momentum conservation" discussed by Scheidl and Vinokur\cite{Vinokur}.

The differential equation $ M \ddot{y} - \eta \dot{y} = \rho'-\rho $ with boundary conditions $y(0)=y_0, y(t)=0$ is solved as
\begin{align}
y(t')&=y_h(t') + y_p(t') \nonumber \\
y_h(t')&=\frac{y_0}{\gamma d(t)}\left[1-e^{-\gamma(t-t')}\right] \nonumber \\
y_p(t')&=-\frac{1}{\eta} \left[ \sum_{i=1}^n k_i h(t'-t_i)+\sum_{i=1}^m k'_i h(t'-t'_i) \right]
\end{align}
with $\gamma=\eta /M$, and $h(t') = e^{\gamma t'} \Theta(-t') + \Theta(t')$.

\subsection{S3-2. Mobility}
As shown by Fisher and Zwerger, the nonlinear mobility of the system is 
\begin{equation}
\frac{\mu}{\mu_0} = 1-\lim_{t\rightarrow\infty} \frac{1}{Ft} \Braket{\int_0^t \td t' \frac{1}{2} (\rho + \rho') }.
\end{equation}
The average is defined as
\begin{align}
\Braket{A}&=\sum_{n,m=0}^\infty  (-i)^n i^m \int_0^t \td t_1\cdots\int_0^{t_{n-1}} \td t_n  \int_0^t \td t'_1 \cdots\int_0^{t'_{m-1}} \td t'_m \sum_{k_1\cdots k_n,k_1'\cdots k'_n} \left(\prod_{i=1}^n V_{k_i} \right)  \left( \prod_{i=1}^m V_{k'_i} \right)  A \exp \Omega [y_p]  \\
\Omega&= i\int_0^t \td t'\  \left[ F  -\frac{1}{2}(\rho + \rho')  \right] y_p(t') - S_2[y_p] \label{omega} 
\end{align}

\subsection{S3-3. Duality mapping}
As shown by Fisher and Zwerger, charge densities in original model are transcripted as sharp tight-binding trajectories;
\begin{equation}
q_s(t') = -\frac{1}{\eta}\sum_{i=1} ^n k_i \theta (t'-t_i),\qquad 	q'_s(t') = \frac{1}{\eta} \sum_{i=1} ^m k'_i \theta (t'-t_i)
\end{equation}
In terms of $x_s =(q_s+q_s')/2 $ and $y_s=q_s-q'_s$,
\begin{equation}
x_s(t') = \frac{1}{\eta} \int_0^{t'} \td t'' \frac{1}{2} (\rho+\rho'), \qquad y_s(t') = \frac{1}{\eta} \int_0^{t'} \td t''  (\rho-\rho').
\end{equation}
They are evaluated as
\begin{align}
x_s(t') &= \frac{1}{2\eta} \int_0^{t'} \td t'' \left[ -\sum_i k_i \delta(t''-t_i) +\sum_i k_i'\delta(t''-t_i')\right] = \frac{1}{2\eta}\left[ -\sum_i k_i \Theta(t'-t_i) +\sum_i k_i'\Theta(t'-t_i')\right]  \\
\de{x_s}{t'}&= \frac{1}{2\eta}\left[ -\sum_i k_i \delta(t'-t_i) +\sum_i k_i'\delta(t'-t_i')\right] 
\end{align}
At the boundary of its domain,
\begin{equation}
x_s(t'\leq 0) = 0,\qquad x_s(t'\geq t) = \frac{1}{2\eta}\left[ -\sum_i k_i  +\sum_i k_i'\right]  = \frac{1}{\eta} \sum_i k_i', \qquad \de{x_s}{t'}(t'\leq 0)= \de{x_s}{t'}(t' \geq t)=0 
\end{equation}
For $y_s$ coordinate,
\begin{align}
y_s(t') &= \frac{1}{\eta} \int_0^{t'} \td t'' \left[ -\sum_i k_i \delta(t''-t_i) -\sum_i k_i'\delta(t''-t_i')\right] = -\frac{1}{\eta}\left[ \sum_i k_i \Theta(t'-t_i) +\sum_i k_i'\Theta(t'-t_i')\right]  \\
\de{y_s}{t'}&= -\frac{1}{\eta}\left[ \sum_i k_i \delta(t'-t_i) +\sum_i k_i'\delta(t'-t_i')\right] 
\end{align}
and
\begin{equation}
y_s(t'\leq 0) = 0,\qquad y_s(t'\geq t) = \frac{1}{\eta}\left[ -\sum_i k_i  -\sum_i k_i'\right]  =0,  \qquad \de{y_s}{t'}(t' \leq 0)= \de{y_s}{t'}(t'\geq t)=0 
\end{equation}

The weight in the average in the tight binding picture is calculated as eq. (\ref{omega}). 
The first term is
\begin{align}
iF\int_0^t \td t'\  y_p(t') &=  -iF\frac {1}{\eta} \int_0^t \td t'\  \left[ \sum_{i=1}^n k_i h(t'-t_i)+\sum_{i=1}^m k'_i h(t'-t'_i) \right]\nonumber \\
&=-iF\frac {1}{\eta}  \left[ \sum_{i=1}^n k_i  \left( \frac{1}{\gamma}(1-e^{-\gamma t_i }) + t-t_i \right)+\sum_{i=1}^m k'_i \left( \frac{1}{\gamma}(1-e^{-\gamma t'_i }) + t-t'_i \right) \right]\nonumber \\
&\simeq iF\frac {1}{\eta}  \left[ \sum_{i=1}^n k_i  t_i +\sum_{i=1}^m k'_i t'_i  \right] = iF\int_0^t \td t'\  y_s(t') 
\end{align}
where we have neglected exponentially small boundary terms.

Before going further, we see the particular trajectory $y_p$ is expressed in terms of the sharp tight binding trajectory $y_s$ as
\begin{equation}
M \ddot{y_p} - \eta \dot{y_p} = \rho'-\rho = -\eta \dot{y}_s \qquad \Rightarrow \qquad  y_p(\omega) = \frac{\gamma}{\gamma+i\omega} y_s(\omega) .
\end{equation} 
Inserting this relation to eq. (\ref{omega}), and using $\de{x_s}{t'}=y_s(t')= 0 \ (t'\notin[0,t])$, the last two terms in large $t$ limit reads
\begin{align}
-i\int_0^t \td t'\  \frac{1}{2}(\rho + \rho') y_p(t') - S_2[y_p]  = i\Phi^{(\gamma)} (x_s,y_s)
\end{align}
where $\Phi^{(\gamma)}$ is evaluated in eq.(\ref{Phi}) using the modified spectral function of the bath: 
\begin{equation}\label{Jomega}
J^{(\gamma)}(\omega) =\frac{\eta \omega}{1+(\omega/\gamma)^2} 
\end{equation}
instead of $J(\omega) = \eta \omega$. 

Using them, the mobility is calculated as
\begin{equation}
\frac{\mu}{\mu_0} = 1-\frac{1}{\mu_0}\lim_{t\rightarrow\infty} \frac{\Braket{x_s(t)}}{Ft} .
\end{equation}

\subsection{S3-4. Evaluation of $\Omega$}
In this section, we calculate each term in eq.(\ref{omega}).  We omit the superscript $(\gamma)$ to simplify the notations. 

\subsubsection{S3-4-1. First term in $\Omega$}
\begin{align}
iF \int_0^t \td t' y_s(t') &=-\frac{iF}{\eta} \int_0^t  \td t'\left[ \sum_i k_i \Theta(t'-t_i) +\sum_i k_i'\Theta(t'-t_i')\right]   \nonumber \\
&=-\frac{iF}{\eta} \left[ \sum_i k_i (t-t_i) +\sum_i k_i'(t-t_i')\right] =\frac{iF}{\eta} \left[ \sum_i k_i t_i +\sum_i k_i't_i'\right]  
\end{align}

\subsubsection{S3-4-2. Evaluation of $\Phi^{(\gamma)}$} 
Before the evaluation of the influence phase $\Phi$, we define some new functions;
\begin{align}
\alpha_I(s-t') &= -\int_0^\infty \frac{\td \omega}{\pi } J(\omega) \sin\omega (s-t') \nonumber \\
&= -\der{s} \der{t'} \int_0^\infty \frac{\td \omega}{\pi \omega^2} J(\omega) \sin\omega (s-t')  \equiv -\der{s} \der{t'}\frac{\eta}{\pi} Q_1(s-t') \\
\alpha_R(t'-s) &= \int_0^\infty \frac{\td \omega}{\pi } J(\omega) \cos\omega (t'-s) \coth\left(\frac{\hbar \omega}{2T}\right) \nonumber \\
&=-\der{s} \der{t'} \int_0^\infty \frac{\td \omega}{\pi \omega^2 } J(\omega) \left[1-\cos\omega (t'-s)\right] \coth\left(\frac{\hbar \omega}{2T}\right) \equiv   \der{s} \der{t'}\frac{\eta}{\pi} Q_2(t'-s).
\end{align}

Now we calculate $\Phi$. The first two terms are
\begin{align}
& -i\int_0^t \td t' \int_{t'}^t \td s \ 2 x_s(t')\alpha_I(s-t') y_s(s) -i M (\Delta \omega)^2 \int_0^t \td t'  x_s(t')y_s(t') \nonumber \\
&=i\frac{\eta}{\pi}  \int_0^t \td t' \int_{t'}^t \td s \ 2 x_s(t')  y_s(s) \der{s}\der{t'} Q_1(s-t') -i M (\Delta \omega)^2 \int_0^t \td t'  x_s(t')y_s(t') \nonumber \\
&=  i\frac{2\eta}{\pi} \int_0^t \td t' \int_{t'}^t \td s \de{x_s}{t'} \de{y_s}{s} Q_1(s-t') \nonumber \\
&= -i\frac{1}{\pi\eta} \sum_{i,j} \left( - k_i \left[ k_j \Theta(t_j-t_i)Q_1(t_j-t_i) +k_j'\Theta(t_j'-t_i)Q_1(t_j'-t_i)\right] \right.  \nonumber \\
& \qquad \qquad \qquad \qquad \left. + k_i'\left[ k_j \Theta(t_j-t'_i)Q_1(t_j-t'_i) +k_j'\Theta(t_j'-t'_i)Q_1(t_j'-t'_i)\right]\right).
\end{align}
Similarly, the last term is
\begin{align}
& - \int_0^t \td t' \int_0^{t'} \td s \ y_s(t')\alpha_R(t'-s) y_s(s) \nonumber \\
&= \frac{1}{\pi\eta} \sum_{i,j} \left(  k_i \left[ k_j \Theta(t_i-t_j)Q_2(t_i-t_j) +k_j'\Theta(t_i-t'_j)Q_2(t_i-t'_j)\right] \right.  \nonumber \\
& \qquad \qquad \qquad \qquad \left. + k_i'\left[ k_j \Theta(t'_i-t_j)Q_2(t'_i-t_j) +k_j'\Theta(t_i'-t'_j)Q_2(t_i'-t'_j).\right]\right)
\end{align}

\subsubsection{S3-4-3. General expression}

Thus, we finally have the explicit expression for the steady velocity in arbitrary order of $V$'s;
\begin{align}
\frac{\mu}{\mu_0} &= 1-\frac{1}{\mu_0}\lim_{t\rightarrow\infty} \frac{\Braket{x_s(t)}}{Ft} \\
\Braket{x_s(t)}&=\sum_{n,m=0}^\infty  (-i)^n i^m \int_0^t \td t_1\cdots\int_0^{t_{n-1}} \td t_n  \int_0^t \td t'_1 \cdots\int_0^{t'_{m-1}} \td t'_m \sum_{k_1\cdots k_n,k_1'\cdots k'_n} \left(\prod_{i=1}^n V_{k_i} \right)  \left( \prod_{i=1}^m V_{k'_i} \right)  \frac{1}{\eta} \left( \sum_i k_i'\right)  \exp \Omega [y_s] \\
\Omega&=\frac{iF}{\eta} \sum_i \left[  k_i t_i +k_i't_i'\right]   -i\frac{1}{\pi\eta} \sum_{i,j} \left( - k_i \left[ k_j \Theta(t_j-t_i)Q_1(t_j-t_i) +k_j'\Theta(t_j'-t_i)Q_1(t_j'-t_i)\right] \right.  \nonumber \\
& \qquad \qquad \qquad \qquad \qquad \qquad \qquad \qquad \left. + k_i'\left[ k_j \Theta(t_j-t'_i)Q_1(t_j-t'_i) +k_j'\Theta(t_j'-t'_i)Q_1(t_j'-t'_i)\right]\right) \nonumber \\
& \qquad \qquad \qquad \qquad \qquad +\frac{1}{\pi\eta} \sum_{i,j} \left(  k_i \left[ k_j \Theta(t_i-t_j)Q_2(t_i-t_j) +k_j'\Theta(t_i-t'_j)Q_2(t_i-t'_j)\right] \right.  \nonumber \\
& \qquad \qquad \qquad \qquad \qquad \qquad \qquad \qquad \left. + k_i'\left[ k_j \Theta(t'_i-t_j)Q_2(t'_i-t_j) +k_j'\Theta(t_i'-t'_j)Q_2(t_i'-t'_j).\right]\right)
\end{align}

\subsection{S3-5. Order of $V^2$}
In the order of $V^2$, only the contribution comes from the configuration with $m=n=1$. For $m=n=1$, the calculation is now
\begin{align}
\Braket{x_s(t)}&= (-i) i \int_0^t \td t_1  \int_0^t \td t'_1 \sum_{k_1,k_1',k_1+k_1'=0} V_{k_1} V_{k'_1}  \frac{1}{\eta} k_i'  \exp \Omega [y_s] \\
\Omega&=\frac{iF}{\eta} \left[  k_1 t_1 +k_1't_1'\right]  -i\frac{1}{\pi\eta} \left( - k_1 \left[ k_1'\Theta(t_1'-t_1)Q_1(t_1'-t_1)\right]  + k_1'\left[ k_1 \Theta(t_1-t'_1)Q_1(t_1-t'_1) \right]\right) \nonumber \\
&\qquad\qquad\qquad\qquad \qquad+\frac{1}{\pi\eta}  \left(  k_1 \left[ k_1'\Theta(t_1-t'_1)Q_2(t_1-t'_1)\right]  + k_1'\left[ k_1 \Theta(t'_1-t_1)Q_2(t'_1-t_1)\right]\right) \nonumber \\
&=\frac{iF}{\eta} \left[  k_1 t_1 +k_1't_1'\right]    +i\frac{1}{\pi\eta}  k_1  k_1' Q_1(t_1-t'_1)  +\frac{1}{\pi\eta}    k_1  k_1'Q_2(t_1-t'_1) 
\end{align}
Here we have used $Q_1(-t')=-Q_1(t')$, $Q_2(-t')=Q_2(t')$ and $Q_1(0)=Q_2(0)=0$. Symmetrizing in $k$ summation, we have
\begin{align}
\Braket{x_s(t)}&=  -\frac{1}{\eta}  \int_0^t \td t_1  \int_0^t \td t'_1 \sum_{k} \left| V_{k} \right|^2 k  \exp \left[\frac{iF}{\eta} k\left(t_1-t_1'\right)    -i\frac{1}{\pi\eta} k ^2Q_1(t_1-t'_1)  -\frac{1}{\pi\eta}     k^2Q_2(t_1-t'_1) \right] \nonumber \\
&=  -\frac{i}{\eta}  \int_0^t \td t_1  \int_0^t \td t'_1 \sum_{k} k \left| V_{k} \right|^2  \sin\left[\frac{F}{\eta} k\left(t_1-t_1'\right)   \right] \exp \left[ -i\frac{1}{\pi\eta} k ^2Q_1(t_1-t'_1)  -\frac{1}{\pi\eta}     k^2Q_2(t_1-t'_1) \right] 
\end{align}
As we see $ \int_0^t \td t_1  \int_0^t \td t'_1 f(t_1-t'_1) \rightarrow    t \int_{0}^{\infty} \td \Delta \left[ f(\Delta)+f(-\Delta) \right] 	$ in the long time limit, the velocity reads
\begin{align} \label{Vinokurv2}
\lim_{t\rightarrow\infty} \frac{\Braket{x_s(t)}}{t}&=  -\frac{i}{\eta}  \int_0^\infty \td \Delta \sum_{k} k \left| V_{k} \right|^2 \left(   \sin\left[\frac{F}{\eta} k\Delta   \right] \exp \left[ -i\frac{1}{\pi\eta} k ^2Q_1(\Delta)  -\frac{1}{\pi\eta}     k^2Q_2(\Delta) \right] +(\Delta\rightarrow -\Delta)\right)\nonumber \\
&=  \frac{2}{\eta}  \int_0^\infty \td \Delta \sum_{k} k \left| V_{k} \right|^2   \sin\left[\frac{F}{\eta} k \Delta  \right] \sin \left[ \frac{1}{\pi\eta} k ^2Q_1(\Delta) \right] \exp\left[  -\frac{1}{\pi\eta}   k^2Q_2(\Delta) \right].
\end{align}
This is the eq.(1) in the main text. This result is the same as Peguiron-Grifoni's result (eq.(9),(10)) \cite{Peguiron,Peguiron2} and reduce to the Scheidl-Vinokur's result (eq.(53)) \cite{Vinokur} in the small $F$ limit and to Fisher-Zwerger's result (eq.(3.51),(4.2)) \cite{Fisher} if we take only $k=\pm\frac{2\pi}{a}$.

Note that we used the exponential cutoff for the bath spectral function $J(\omega)=\eta \omega e^{-\omega/\gamma}$ in the main text instead of the Lorentzian cutoff (eq.(\ref{Jomega})).

\subsection{S3-6. Order of $V^3$}
In this case, two possibilitis $(m,n) = (2,1),(1,2)$ are allowed.
\begin{align}
\Braket{x_s(t)}&= -\frac{i}{\eta} \int_0^t \td t_1\int_0^{t_1} \td t_2  \int_0^t \td t'_1 \sum_{k_1+k_2+k_1'=0} V_{k_1}V_{k_2} V_{k'_1}  k'_1 \nonumber \\
&\qquad  \times \exp\left[ \frac{iF}{\eta} \left[  k_1 t_1 +k_2t_2 + k_1't_1'\right] \right. \nonumber \\
&\qquad   -i\frac{1}{\pi\eta}  \left( - k_1 \left[ k_2 \Theta(t_2-t_1)Q_1(t_2-t_1) +k_1'\Theta(t_1'-t_1)Q_1(t_1'-t_1)\right]  \right.\nonumber \\
&\qquad  \qquad \qquad - k_2 \left[ k_1 \Theta(t_1-t_2)Q_1(t_1-t_2) +k_1'\Theta(t_1'-t_2)Q_1(t_1'-t_2)\right] \nonumber   \\
&\qquad  \qquad \qquad \left.+ k_1'\left[ k_1 \Theta(t_1-t'_1)Q_1(t_1-t'_1)+ k_2 \Theta(t_2-t'_1)Q_1(t_2-t'_1) \right]\right) \nonumber \\
&\qquad  +\frac{1}{\pi\eta} \left(  k_1 \left[ k_2 \Theta(t_1-t_2)Q_2(t_1-t_2) +k_1'\Theta(t_1-t'_1)Q_2(t_1-t'_1)\right] \right.  \nonumber \\
&\qquad  \qquad \qquad +k_2\left[ k_1 \Theta(t_2-t_1)Q_2(t_2-t_1) +k_1'\Theta(t_2-t'_1)Q_2(t_2-t'_1)\right]  \nonumber \\
&\qquad  \qquad \qquad \left. \left. + k_1'\left[ k_1 \Theta(t'_1-t_1)Q_2(t'_1-t_1) +k_2 \Theta(t'_1-t_2)Q_2(t'_1-t_2) \right]\right) \right] \nonumber \\
& +\frac{i}{\eta} \int_0^t \td t_1  \int_0^t \td t'_1\int_0^{t'_1} \td t'_2 \sum_{k_1+k_1'+k'_2=0} V_{k_1} V_{k'_1} V_{k'_2}  (k'_1+k'_2)  \nonumber \\
&\qquad  \times\exp\left[ \frac{iF}{\eta} \left[  k_1 t_1+ k_1't_1' +k'_2t'_2 \right]  \right. \nonumber \\
& \qquad  -i\frac{1}{\pi\eta}  \left( - k_1 \left[ k_1'\Theta(t_1'-t_1)Q_1(t_1'-t_1)+k_2'\Theta(t_2'-t_1)Q_1(t_2'-t_1)\right] \right.  \nonumber \\
& \qquad \qquad \qquad  + k_1'\left[ k_1 \Theta(t_1-t'_1)Q_1(t_1-t'_1) +k_2'\Theta(t_2'-t'_1)Q_1(t_2'-t'_1)\right] \nonumber \\
& \qquad \qquad \qquad \left.+ k_2'\left[ k_1 \Theta(t_1-t'_2)Q_1(t_1-t'_2) +k_1'\Theta(t_1'-t'_2)Q_1(t_1'-t'_2)\right]\right) \nonumber \\
&\qquad  +\frac{1}{\pi\eta} \left(  k_1 \left[ k_1'\Theta(t_1-t'_1)Q_2(t_1-t'_1)+ k_2'\Theta(t_1-t'_2)Q_2(t_1-t'_2) \right] \right.  \nonumber \\
&\qquad  \qquad \qquad  + k_1'\left[ k_1 \Theta(t'_1-t_1)Q_2(t'_1-t_1) +k_2'\Theta(t_1'-t'_2)Q_2(t_1'-t'_2)\right] \nonumber \\
&\qquad  \qquad \qquad \left. \left. + k_2'\left[ k_1 \Theta(t'_2-t_1)Q_2(t'_2-t_1) +k_1'\Theta(t_2'-t'_1)Q_2(t_2'-t'_1)\right]\right) \right] \nonumber \\
\end{align}
\begin{align}
&= -\frac{i}{\eta} \int_0^t \td t_1\int_0^{t_1} \td t_2  \int_0^t \td t'_1 \sum_{k_1+k_2+k_1'=0} V_{k_1}V_{k_2} V_{k'_1}  k'_1  \nonumber\\
&\qquad    \times\exp \left[\frac{iF}{\eta} \left[  k_1 t_1 +k_2t_2 + k_1't_1'\right] \right. \nonumber \\
&\qquad   -i\frac{1}{\pi\eta}  \left(   - k_1 k_2 Q_1(t_1-t_2) -  k_1 k_1'Q_1(t'_1-t_1)- k_2k_1' Q_1(t'_1-t_2) \right) \nonumber \\
&\qquad  \left. +\frac{1}{\pi\eta} \left(  k_1 k_2 Q_2(t_1-t_2) +k_1k_1'Q_2(t_1-t'_1) + k_1' k_2 Q_2(t'_1-t_2) \right) \right] \nonumber \\
&+\frac{i}{\eta} \int_0^t \td t_1  \int_0^t \td t'_1\int_0^{t'_1} \td t'_2 \sum_{k_1+k_1'+k'_2=0} V_{k_1} V_{k'_1} V_{k'_2}  (k'_1+k'_2 )\nonumber \\
&\qquad   \times\exp \left[\frac{iF}{\eta} \left[  k_1 t_1+ k_1't_1' +k'_2t'_2 \right]  \right. \nonumber \\
&\qquad   -i\frac{1}{\pi\eta}  \left( - k_1 k_1'Q_1(t_1'-t_1)-k_1 k_2'Q_1(t_2'-t_1) - k_1'k_2'Q_1(t_2'-t'_1)\right) \nonumber \\
&\qquad  \left. +\frac{1}{\pi\eta} \left(  k_1' k_1 Q_2(t'_1-t_1) +k'_1k_2'Q_2(t_1'-t'_2) + k_2'k_1 Q_2(t'_2-t_1)\right) \right]
\end{align}
By exchanging variables with and without prime(') in the second term we get
\begin{align}
\Braket{x_s(t)}&= -\frac{2i}{\eta} \int_0^t \td t_1\int_0^{t_1} \td t_2  \int_0^t \td t'_1 \sum_{k_1+k_2+k_1'=0} V_{k_1}V_{k_2} V_{k'_1}  k'_1  \nonumber\\
&\qquad \times\exp\left[  \frac{iF}{\eta} \left[  k_1 t_1 +k_2t_2 + k_1't_1'\right]  +\frac{1}{\pi\eta} \left(  k_1 k_2 Q_2(t_1-t_2) +k_1k_1'Q_2(t_1-t'_1) + k_1' k_2 Q_2(t'_1-t_2) \right) \right] \nonumber \\
&\qquad \times\cos\left[ \frac{1}{\pi\eta}  \left(  k_1 k_2 Q_1(t_1-t_2) +  k_1 k_1'Q_1(t'_1-t_1) + k_2k_1' Q_1(t'_1-t_2) \right) \right]
\end{align}
The velocity should be real therefore we can take the real part. In the long time limit, using $ \int_0^t \td t_1\int_0^{t_1} \td t_2  \int_0^t \td t'_1 f(t_1-t_2,t_2-t_1')  \rightarrow t \int_0^{\infty} \td \Delta  \int_{0}^{\infty} \td s \left[f(\Delta,s) +f(\Delta,-s) \right]$, we get
\begin{align}
\lim_{t\rightarrow\infty} \frac{\Braket{x_s(t)}}{t}
&= \frac{2}{\eta}  \int_0^{\infty} \td \Delta  \int_{0}^{\infty} \td s\sum_{k_1+k_2+k_1'=0}  k'_1 \nonumber \\
&\qquad \times\left( \Re\left[  V_{k_1}V_{k_2} V_{k'_1} \right]\sin\left[ \frac{F}{\eta} \left[  k_1 \Delta - k_1' s\right] \right]+  \Im\left[  V_{k_1}V_{k_2} V_{k'_1} \right] \cos\left[ \frac{F}{\eta} \left[  k_1 \Delta - k_1' s\right] \right]\right) \nonumber\\
&\qquad \times \exp\left[ \frac{1}{\pi\eta} \left(  k_1 k_2 Q_2(\Delta) +k_1k_1'Q_2(\Delta+s) + k_1' k_2 Q_2(s) \right) \right] \nonumber \\
&\qquad  \times\cos\left[ \frac{1}{\pi\eta}  \left( k_1 k_2 Q_1(\Delta) -  k_1 k_1'Q_1(\Delta+s) - k_2k_1' Q_1(s) \right) \right]\nonumber \\
& \qquad\qquad  +(s\rightarrow -s)
\end{align}
Since $\int_0^{\infty} \td \Delta \int_0^{\infty} \td s \left[ f(\Delta,s)+f(\Delta,-s)\right] = \int_0^{\infty} \td \Delta \int_0^{\infty} \td s \left[ f(\Delta,s)+f(\Delta+s,-s)+f(s,-\Delta-s)\right]$, the result is rewritten after some calculations by using $Q_1(-t) = -Q_1(t)$, $Q_2(-t) = Q_2(t)$ and the momentum conservation followed by relabeling variables as
\begin{align} \label{Vinokurv3}
\lim_{t\rightarrow\infty} \frac{\Braket{x_s(t)}}{t}
&=- \frac{4}{\eta}  \int_0^{\infty} \td \Delta  \int_{0}^{\infty} \td s\sum_{k_1+k_2+k_1'=0} k_1 \nonumber \\
& \qquad \times\left( \Re\left[  V_{k_1}V_{k_2} V_{k'_1} \right] \sin\left[ \frac{F}{\eta} \left(  k_1 \Delta - k_1' s\right) \right]+\Im\left[  V_{k_1}V_{k_2} V_{k'_1} \right]  \cos\left[ \frac{F}{\eta} \left(  k_1 \Delta - k_1' s\right) \right]\right) \nonumber\\
&\qquad \times    \exp\left[ \frac{1}{\pi\eta} \left(  k_1 k_2 Q_2(\Delta) +k_1k_1'Q_2(\Delta+s) + k_1' k_2 Q_2(s) \right) \right]\nonumber \\
&\qquad \times\sin \left[ \frac{1}{\pi\eta}  k_1 k_2 Q_1(\Delta)\right]  \sin \left[ \frac{1}{\pi\eta} \left(  k_1 k_1'Q_1(\Delta+s) + k_2k_1' Q_1(s)\right)  \right] .
\end{align}
After the subtraction of the value at $F=0$, we get the eq.(8) in the main text. 
This is consistent with the Scheidl-Vinokur's result \cite{Vinokur} for the order of $F^2$ and the Peguiron-Grifoni's result for the rectified velocity $v(F)+v(-F)$ in the presence of up to the second harmonic potential; $k=\pm\frac{2\pi}{a},\pm\frac{4\pi}{a}$  \cite{Peguiron,Peguiron2}.

\section{S4. Another derivation of eq.(1) and (8)}

In this section, we show another derivation of eq.(1) and eq.(8) following the formalisms developed by Eckern and Pelzer \cite{Eckern,Schmid2}

\subsection{S4-1. Setup}
In the Keldysh formalism, the time integral is composed of to paths; $C=C_1 +C_2$ where $C_1$ runs from $t=-\infty$ to $t=\infty$ and  $C_2$ runs from $t=\infty$ to $t=-\infty$. They are labeled as $\beta=1,2$, respectively. Accordingly, there are in general two operators $q_1(t),q_2(t)$ for given time $t$, summarized as $\vec{q} = (q_1,q_2)$. It is convenient to introduce $x_1 = (q_1-q_2)/\sqrt{2}$, and $x_2 = (q_+-q_2)/\sqrt{2}$.
The action is
\begin{align}
S[\vec{q}] &= S_0  + S_{p} + S_F \\
S_{p} &=-\int_C \td t    V(q(t)) = -\int_{-\infty}^{\infty} \td t    V(q_1(t))-V(q_2(t)) \\
S_{F}  &= -F\int_{-\infty}^{\infty} \td t    \left( q_1(t)-q_2(t) \right) = -\sqrt{2} F\int_{-\infty}^{\infty} \td t    x_1(t) 
\end{align}
$S_0$ expresses inertia and Ohmic dissipation whose form is specified later. We define the generating functional
\begin{equation}
Z[\xi] = \Braket{  T_C \exp\left[ i \int \td t \  \vec{\xi} (t)  \cdot \vec{x}(t) \right]   }
\end{equation}
with contour ordering $T_C$. The expectation value is taken at equilibrium condition. Thus, the differential mobility of the system is 
\begin{equation}
\mu(\omega) =i\omega \cdot \left[ i \left. \frac{\delta^2 Z[\xi]}{\delta \xi_2(t) \delta\xi_1(t')} \right|_{\xi\to 0}\right]_\omega
\end{equation}
$\left[\ \right]_\omega $ means the Fourier transformation with respect to the time difference.

In the perturbation theory, we write the partition function as
\begin{equation}
Z[\xi] = \exp\left[ iS_p\left( -i\pa{\xi}\right)\right] Z_0[\xi]. \label{perturbation}
\end{equation}
$Z_0$, corresponding to $S_0$, is calculated by the standard Gaussian form;
\begin{equation}
Z_0[\xi] = \exp\left[-\frac{i}{2} \Braket{\xi|D_0|\xi}\right]
\end{equation}
with
\begin{equation}
\Braket{\xi|D_0|\xi} = \sum_{\alpha,\beta} \int _{-\infty} ^{\infty} \td t \td t' \ \xi_\alpha(t) D_0^{\alpha\beta}(t,t') \xi_\beta(t').
\end{equation}
The free Keldysh Grren's function is 
\begin{equation}
\underline{D}_0= \left(   \begin{array}{cc} 0 & D_0^A \\ D_0^R & D_0^K\end{array}\right)
\end{equation}
where $D_0^R(\omega)=\left[D_0^A(\omega)\right]^*= (i\eta\omega + m\omega^2)^{-1}$ and $D_0^K(\omega) = \left[D_0^R(\omega) - D_0^A(\omega)\right] \coth(\omega/2T)$ from the fluctuation-dissipation theorem.

The potential term is expressed as
\begin{equation}
S_p = \sum_k V_k \int \td t \left[e^{ikq_1}-e^{ikq_2}\right] = \sum_k \sum_{\delta = \pm 1} V_k \delta \int \td t e^{ikx_2/\sqrt{2}}e^{i\delta k x_1/\sqrt{2}} .
\end{equation}
The exponentiated one is expanded as
\begin{align}
e^{iS_p} &= \sum_{n=0}^{\infty } \frac{(-i)^n}{n!}\sum_{k_1\dots k_n} \sum_{\delta_1\dots \delta_n}\delta_1\dots \delta_n V_{k_1}\dots V_{k_n} \int \td t_1\dots\td t_n \exp\left[ \int_{-\infty}^{\infty} \td t \   \vec{\rho}^{(n)} (t) \cdot \vec{x} (t)\right] \nonumber \\
&=: \mathrm{Tr} \exp\left[ i \Braket{\rho^{(n)}|x} \right]
\end{align}
with
\begin{equation}
\vec{\rho}^{(n)} (t) = \frac{1}{\sqrt{2}} \sum_{i=1}^n  \left(   \begin{array}{c} \delta_i  \\ 1 \end{array}\right) k_i\delta(t-t_i).
\end{equation}
Here we define the "$\mathrm{Tr}$" in the above expression. Inserting this expression to eq.(\ref{perturbation}), we can write
\begin{equation}
Z[\xi] = \mathrm{Tr} Z_0[\xi+\rho]. \label{partition}
\end{equation}
Here, the normalization is preserved in this expression as $Z[0] = \mathrm{Tr}Z_0[\rho]=1$ as following. The summation over $\delta$ gives us the factor $\prod_j \sin \left[ \sum_i \frac{k_ik_j}{2} D_0^R(t_i-t_j) \right]$, which vanishes due to the causality; $D_0^R(t<0) =0$. Only the remaining term is for $n=0$ which gives $\mathrm{Tr}Z_0[\rho]=1$.

Perturbed connected Green's function is calculated as
\begin{equation}
D^{\alpha\beta} (t,t') = \left. i \frac{\delta^2 \log Z[\xi]}{\delta\xi_\alpha(t) \delta \xi_\beta(t') } \right|_{\xi\to 0}.
\end{equation}
Directly Inserting eq.(\ref{partition}), we find 
\begin{equation}
\underline{D} (\omega)=  \underline{D}_0(\omega) \underline{S} (\omega) \underline{D}_0 (\omega)
\end{equation}
with
\begin{equation}
S^{\alpha\beta}(t,t') = -i\Braket{\rho_\alpha(t)\rho_\beta(t')} \equiv -i \mathrm{Tr} \left[ Z_0[\rho] \rho_\alpha(t)\rho_\beta(t') \right] .
\end{equation}
One can easily find that $S_{22} = 0 $ by $\delta$-summation. 
The self-energy is computed as
\begin{equation}
\Sigma = D_0^{-1} -D^{-1} = D_0^{-1} \left[1-(1+D_0 S)^{-1}\right]= S(1+D_0 S)^{-1}.
\end{equation}
Since the mobility is expressed in terms of retarded component of the Green's function, the important relation is
\begin{equation}
\Sigma^R = (1+S_{12}D_0^A)^{-1} S_{12}.
\end{equation}
Once the retarded self-energy is obtained, the mobility is
\begin{equation}
\mu(\omega) = \frac{i\omega}{i\eta \omega + m\omega^2 + \Sigma^R(\omega) } .
\end{equation}
Especially in the DC limit,
\begin{equation}
\mu(\omega = 0) = \frac{1}{i\eta^2} \left. \p{ \Sigma^R(\omega)}{\omega}\right|_{\omega\to 0}  .
\end{equation}

The force $F$ is introduced as $\xi\to\xi+\sqrt{2}F$ in the generating functional. Equivalently, we put  $\rho\to\rho+\sqrt{2}F$ in $Z_0[\rho]$ in the following. In this substitution, the force factor appears as
\begin{align}
Z_0[\rho] &= \exp\left[ -\frac{i}{2}\int \td t \int \td t' \ \left(\rho_1(t)+\sqrt{2}F\right)D_0^A(t-t')\rho_2(t') \right. \nonumber \\
&\qquad\qquad\qquad\qquad \left.  + \rho_2(t) D_0^R(t-t')\left(\rho_1(t')+\sqrt{2}F\right) + \rho_2(t) D_0^K(t-t')\rho_2(t')\right] \nonumber  \\
&=\left. Z_0[\rho] \right|_{F=0} \mathcal{P}_F \\
\mathcal{P}_F& =\exp\left[ -\frac{iF}{\sqrt{2}}\int \td t \int \td t' \   D_0^A(t-t')\rho_2(t') + \rho_2(t) D_0^R(t-t')   \right] = \exp\left[ \frac{iF}{\eta}  \sum_{i} k_i t_i \right]
\end{align}
where we have used $D_0^R(t)=D_0^A(-t) = -\Theta(t) (1-e^{-\gamma t})/\eta$ and the momentum conservation $\sum_ik_i=0$ discussed later.

\subsection{S4-2. Order of $V$}
For the first order in $V$, as $S_{12}$ is at least order of $V$, the self energy is given as
\begin{equation}
\Sigma^R = S_{12} + o (V^2)
\end{equation}
\begin{align}
S_{12} &=  (-i)(-i) \frac{1}{2}\sum_{k } \sum_{\delta}\delta V_{k} \int \td t_1 \delta k^2 \delta(t-t_1) \delta(t'-t_1)\exp\left[-\frac{i}{4}\left( 2\delta k^2D_0^R(t=0) + k^2D_0^K(t=0) \right)\right] \nonumber \\
&=  -  \delta(t-t') \sum_{k } V_{k} k^2 \exp\left[-\frac{i}{4}  k^2D_0^K(t=0) \right].
\end{align}
Since
\begin{equation}
iD_0^K(t=0) = i\int\td \omega \left[ \frac{1}{i\eta\omega + m\omega^2}-\frac{1}{-i\eta\omega + m\omega^2}\right] \coth\left(\frac{\omega}{2T}\right) =  + \infty,
\end{equation}
the first order contribution vanishes.

For general $n>0$, this divergence imposes a restriction to the momentum configuration. Suppose $t_1=\dots =t_n$, we have;
\begin{align}
\left. Z_0[\rho]\right|_{F=0}&=\exp \left[ -\frac{i}{4}\sum_{ij} k_ik_j(\delta_i,1)  \left(   \begin{array}{cc} 0 & D_0^A \\ D_0^R & D_0^K\end{array}\right)_{(t_i-t_j)} \left(   \begin{array}{c} \delta_j \\ 1 \end{array}\right) \right] \nonumber \\
&\sim \exp \left[ -\frac{i}{4}\sum_{ij} k_ik_j D_0^K(t=0)\right] =\exp \left[ -\frac{i}{4}\left(\sum_{i} k_i \right)^2 D_0^K(t=0)\right] .
\end{align}
Thus, only configurations with $\sum_{i=1}^n k_i = 0$ survive. This is the "momentum conservation" we have seen above.

\subsection{S4-3. Order of $V^2$}
Similar to the case of the first order, as $S_{12}$ is at least order of $V^2$, the self energy is given as
\begin{equation}
\Sigma^R = S_{12} + o (V^3).
\end{equation}
Using causality of Green's functions and the momentum conservation; $k_2=-k_1$,
\begin{align}
Z_0[\rho] &= \exp\left[ -\frac{i}{4}\left[(\delta_1k_1^2D_0^A(0)+\delta_1k_1^2D_0^R(0)+k_1^2D_0^K(0))\right]\right. \nonumber \\
& \qquad \qquad -\frac{i}{4}\left[(\delta_2k_2^2D_0^A(0)+\delta_2k_2^2D_0^R(0)+k_2^2D_0^K(0))\right] \nonumber \\
& \qquad \qquad-\frac{i}{4}\left[(\delta_1k_1k_2D_0^A(t_1-t_2)+\delta_2k_1k_2D_0^R(t_1-t_2)+k_1k_2D_0^K(t_1-t_2))\right]\nonumber \\
&\qquad \qquad\left.  -\frac{i}{4}\left[(\delta_2k_1k_2D_0^A(t_2-t_1)+\delta_1k_1k_2D_0^R(t_2-t_1)+k_1k_2D_0^K(t_2-t_1))\right] \right] \mathcal{P}_F\nonumber \\
&= \exp\left[ \frac{i}{2}k_1^2\left[  \delta_1 D_0^A(t_1-t_2)+\delta_2 D_0^R(t_1-t_2)+D_0^K(t_1-t_2)-D_0^K(0)\right]\right] \mathcal{P}_F 
\end{align}
Therefore we have
\begin{align}
S_{12} &= -i  \frac{-1}{2}\sum_{k_1 k_2} \sum_{\delta_1 \delta_2}\delta_1 \delta_2 V_{k_1} V_{k_2} \int \td t_1\td t_2 \frac{1}{2} \sum_{ij} \delta_ik_ik_j \delta(t-t_i)\delta(t'-t_j) Z_0[\rho] \nonumber \\
&=  \frac{i}{4}\sum_{k_1} \sum_{\delta_1 \delta_2} k_1^2\left| V_{k_1} \right|^2 \int \td t_1\td t_2   \nonumber \\
& \times \left[ \delta_2 \delta(t-t_1)\delta(t'-t_1) +\delta_1 \delta(t-t_2)\delta(t'-t_2) -\delta_2 \delta(t-t_1)\delta(t'-t_2) -\delta_1 \delta(t-t_2)\delta(t'-t_1) \right]  \nonumber \\
&  \times\exp\left[ \frac{i}{2}k_1^2\left[  \delta_1 D_0^A(t_1-t_2)+\delta_2 D_0^R(t_1-t_2)+D_0^K(t_1-t_2)-D_0^K(0)\right]\right] \exp\left[ \frac{iF}{\eta} k_1 (t_1 -t_2) \right] \nonumber\\
&=  \frac{i}{4}\sum_{k_1}k_1^2\left| V_{k_1} \right|^2  \nonumber \\
&\times \left[  \delta(t-t') \int \td t_2  \ 2\cos\left[\frac{k_1^2}{2}D_0^A(t-t_2)\right] 2i\sin\left[\frac{k_1^2}{2}D_0^R(t-t_2)\right] e^{\frac{i}{2}k_1^2\delta D_0^K(t-t_2)}e^{\frac{iF}{\eta} k_1 (t -t_2) } \right.\nonumber \\
& \quad +\delta(t-t') \int \td t_1  \ 2i\sin\left[\frac{k_1^2}{2}D_0^A(t_1-t)\right] 2\cos\left[\frac{k_1^2}{2}D_0^R(t_1-t)\right] e^{\frac{i}{2}k_1^2\delta D_0^K(t_1-t)}e^{\frac{iF}{\eta} k_1 (t_1 -t) } \nonumber \\
& \qquad  \qquad  \qquad  \qquad    -2\cos\left[\frac{k_1^2}{2}D_0^A(t-t')\right] 2i\sin\left[\frac{k_1^2}{2}D_0^R(t-t')\right] e^{\frac{i}{2}k_1^2\delta D_0^K(t-t')}e^{\frac{iF}{\eta} k_1 (t -t') } \nonumber \\
& \qquad  \qquad  \qquad  \qquad  \left.   -2i\sin\left[\frac{k_1^2}{2}D_0^A(t'-t)\right] 2\cos\left[\frac{k_1^2}{2}D_0^R(t'-t)\right] e^{\frac{i}{2}k_1^2\delta D_0^K(t'-t)}e^{\frac{iF}{\eta} k_1 (t' -t) } \right] \nonumber \\
&= -2 \sum_{k_1}k_1^2\left| V_{k_1} \right|^2  \delta(t-t') \int \td t_1 \  \sin\left[\frac{k_1^2}{2}D_0^A(t_1-t)\right] \cos\left[\frac{k_1^2}{2}D_0^R(t_1-t)\right] e^{\frac{i}{2}k_1^2\delta D_0^K(t_1-t)}e^{\frac{iF}{\eta} k_1 (t_1 -t) } \nonumber \\
& +2\sum_{k_1}k_1^2\left| V_{k_1} \right|^2  \cos\left[\frac{k_1^2}{2}D_0^A(t-t')\right] \sin\left[\frac{k_1^2}{2}D_0^R(t-t')\right] e^{\frac{i}{2}k_1^2\delta D_0^K(t-t')}e^{\frac{iF}{\eta} k_1 (t -t') } \nonumber \\
&=  2\sum_{k_1}k_1^2\left| V_{k_1} \right|^2 \sin\left[\frac{k_1^2}{2}D_0^R(t-t')\right] e^{\frac{i}{2}k_1^2\delta D_0^K(t-t')}e^{\frac{iF}{\eta} k_1 (t -t') } -B\delta(t-t')
\end{align}
where we have defined $\delta D_0^K(t) =D_0^K(t)-D_0^K(0)$. The constant $B$ is determined so that $\Sigma^R(\omega=0)=0$ is satisfied.
Finally, we get the expression for the mobility;
\begin{align}
\mu(\omega=0)= \frac{1}{i\eta^2} \int \td t \ it   2\sum_{k_1}k_1^2\left| V_{k_1} \right|^2 \sin\left[\frac{k_1^2}{2}D_0^R(t-t')\right] e^{\frac{i}{2}k_1^2\delta D_0^K(t-t')}e^{\frac{iF}{\eta} k_1 (t -t') }.
\end{align}
Since $D_0^R(t) = -\Theta(t)(1-e^{-\gamma t})/\eta$ and $\delta D_0^K(t) = \frac{2i}{\pi\eta}\int_0^{\infty}\frac{\td \omega}{\omega} \frac{1-\cos\omega t}{1+\omega^2/\gamma^2} \coth\left( \omega/2T\right) \equiv \frac{2i}{\pi\eta} Q_2(t)$ ,
\begin{align}
\mu(\omega=0)= -\frac{2}{\eta^2}  \sum_{k}k^2\left| V_{k} \right|^2  \int_0^{\infty} \td t \  t  \cos\left[ \frac{F}{\eta} kt\right]  \sin\left[\frac{k^2}{2\eta}(1-e^{-\gamma t})\right] e^{-\frac{k^2}{\pi\eta}Q_2(t)} .
\end{align}
The corresponding velocity 
\begin{align}
v=\int_0^F \td F \mu(\omega=0)= -\frac{2}{\eta}  \sum_{k} k \left| V_{k} \right|^2  \int_0^{\infty} \td t \   \sin\left[ \frac{F}{\eta} kt\right]  \sin\left[\frac{k^2}{2\eta}(1-e^{-\gamma t})\right] e^{-\frac{k^2}{\pi\eta}Q_2(t)}
\end{align}
agrees with eq.(\ref{Vinokurv2}) and eq.(1) in the main text.

\subsection{S4-4. Order of $V^3$}
Again, we have
\begin{equation}
\Sigma^R=S_{12} + o (V^4).
\end{equation}

\begin{align}
Z_0[\rho] &= \exp\left[ -\frac{i}{4}k_1^2\left[\delta_1 D_0^A(0)+\delta_1 D_0^R(0)+D_0^K(0)\right]\right. \nonumber \\
& \qquad \qquad -\frac{i}{4}k_2^2\left[ \delta_2D_0^A(0)+\delta_2D_0^R(0)+D_0^K(0)\right] \nonumber \\
& \qquad \qquad -\frac{i}{4}k_3^2\left[ \delta_3D_0^A(0)+\delta_3D_0^R(0)+D_0^K(0)\right] \nonumber \\
& \qquad \qquad-\frac{i}{2}k_1k_2\left[\delta_1D_0^A(t_1-t_2)+\delta_2D_0^R(t_1-t_2)+D_0^K(t_1-t_2)\right]\nonumber \\
& \qquad \qquad-\frac{i}{2}k_2k_3\left[\delta_2D_0^A(t_2-t_3)+\delta_3D_0^R(t_2-t_3)+D_0^K(t_2-t_3)\right]\nonumber \\
& \left. \qquad \qquad-\frac{i}{2}k_3k_1\left[\delta_3D_0^A(t_3-t_1)+\delta_1D_0^R(t_3-t_1)+D_0^K(t_3-t_1)\right]\right]\mathcal{P}_F  \nonumber \\
&= \exp\left[ -\frac{i}{2}k_1k_2\left[\delta_1D_0^A(t_1-t_2)+\delta_2D_0^R(t_1-t_2)+\delta D_0^K(t_1-t_2)\right]\right. \nonumber \\
& \qquad \qquad-\frac{i}{2}k_2k_3\left[\delta_2D_0^A(t_2-t_3)+\delta_3D_0^R(t_2-t_3)+\delta D_0^K(t_2-t_3)\right]\nonumber \\
& \left. \qquad \qquad-\frac{i}{2}k_3k_1\left[\delta_3D_0^A(t_3-t_1)+\delta_1D_0^R(t_3-t_1)+\delta D_0^K(t_3-t_1)\right]\right]\mathcal{P}_F 
\end{align}

\begin{align}
S_{12}(t,t') &= -i\frac{(-i)^3}{3!}\sum_{k_1 k_2 k_3} \sum_{\delta_1 \delta_2 \delta_3}\delta_1 \delta_2 \delta_3 V_{k_1} V_{k_2} V_{k_3} \int \td t_1\td t_2  \td t_3 \frac{1}{2} \sum_{ij} \delta_ik_ik_j \delta(t-t_i)\delta(t'-t_j) Z_0[\rho] \nonumber \\
&= \frac{1}{12}\sum_{k_1 k_2 k_3} V_{k_1} V_{k_2} V_{k_3} \int \td t_1\td t_2  \td t_3  \sum_{j} k_j \delta(t'-t_j) \nonumber \\
& \times\sum_{\delta_1 \delta_2 \delta_3} \left[ k_1\delta_2\delta_3\delta(t-t_1)+ k_2\delta_3\delta_1\delta(t-t_2)+ k_3\delta_1\delta_2\delta(t-t_3)\right] \nonumber \\
& \times\exp\left[ -\frac{i}{2}k_1k_2\left[\delta_1D_0^A(t_1-t_2)+\delta_2D_0^R(t_1-t_2)+\delta D_0^K(t_1-t_2)\right]\right. \nonumber \\
& \qquad \qquad-\frac{i}{2}k_2k_3\left[\delta_2D_0^A(t_2-t_3)+\delta_3D_0^R(t_2-t_3)+\delta D_0^K(t_2-t_3)\right]\nonumber \\
& \left. \qquad \qquad-\frac{i}{2}k_3k_1\left[\delta_3D_0^A(t_3-t_1)+\delta_1D_0^R(t_3-t_1)+\delta D_0^K(t_3-t_1)\right]\right]\mathcal{P}_F \nonumber \\
&= \frac{-8}{12}\sum_{k_1 k_2 k_3} V_{k_1} V_{k_2} V_{k_3} \int \td t_1\td t_2  \td t_3  \sum_{j} k_j \delta(t'-t_j)  \mathcal{P}_F\nonumber \\
& \times\exp\left[ -\frac{i}{2}\left[k_1k_2\delta D_0^K(t_1-t_2)+k_2k_3\delta D_0^K(t_2-t_3)+k_3k_1\delta D_0^K(t_3-t_1)\right]\right] \nonumber \\
& \times\left[ k_1\delta(t-t_1) \cos\left[R_{21}+R_{31}\right] \right. \sin\left[R_{12}+R_{32}\right]\sin\left[R_{23}+R_{13}\right]\nonumber \\
&\qquad  +k_2\delta(t-t_2) \sin\left[R_{21}+R_{31}\right] \cos\left[R_{12}+R_{32}\right]\sin\left[R_{23}+R_{13}\right]\nonumber \\
& \qquad  \left. +k_3\delta(t-t_3) \sin\left[R_{21}+R_{31}\right] \sin\left[R_{12}+R_{32}\right]\cos\left[R_{23}+R_{13}\right] \right]
\end{align}
with $R_{ij} \equiv k_ik_jD_0^R(t_i-t_j)/2$.
\begin{align}
S_{12}(t,t') &= \frac{-2}{3}\sum_{k_1 k_2 k_3} V_{k_1} V_{k_2} V_{k_3} \int \td t_1\td t_2  \td t_3  \mathcal{P}_F\nonumber \\
& \times\exp\left[ -\frac{i}{2}\left[k_1k_2\delta D_0^K(t_1-t_2)+k_2k_3\delta D_0^K(t_2-t_3)+k_3k_1\delta D_0^K(t_3-t_1)\right]\right] \nonumber \\
& \times\left[ \left(3k_1^2\delta(t-t_1)\delta(t'-t_1) + 2k_1k_2\delta(t-t_1)\delta(t'-t_2)\right) \cos\left[R_{21}+R_{31}\right] \right. \sin\left[R_{12}+R_{32}\right]\sin\left[R_{23}+R_{13}\right]\nonumber \\
&\qquad \qquad \qquad \qquad\qquad   +2k_2k_3\delta(t-t_2)\delta(t'-t_3) \sin\left[R_{21}+R_{31}\right] \cos\left[R_{12}+R_{32}\right]\sin\left[R_{23}+R_{13}\right]\nonumber \\
& \qquad \qquad \qquad \qquad \qquad  \left. +2k_3k_1\delta(t-t_3)\delta(t'-t_1) \sin\left[R_{21}+R_{31}\right] \sin\left[R_{12}+R_{32}\right]\cos\left[R_{23}+R_{13}\right] \right] \nonumber \\
& =-4\sum_{k_1 k_2 k_3} V_{k_1} V_{k_2} V_{k_3} \int \td t_1 \exp\left[\frac{iF}{\eta}(k_1t_1+k_2t+k_3t')\right]\nonumber \\
& \times\exp\left[ -\frac{i}{2}\left[k_1k_2\delta D_0^K(t_1-t)+k_2k_3\delta D_0^K(t-t')+k_3k_1\delta D_0^K(t'-t_1)\right]\right] \nonumber \\
& \times  k_2k_3\cos\left[\frac{k_1k_2}{2}D_0^R(t_1-t)\right] \sin\left[ \frac{k_2k_3}{2}D_0^R(t-t')+\frac{k_3k_1}{2}D_0^R(t_1-t') \right]\sin\left[ \frac{k_1k_2}{2}D_0^R(t-t_1)+\frac{k_3k_1}{2}D_0^R(t'-t_1) \right]\nonumber \\
& -2\delta(t-t')\sum_{k_1 k_2 k_3} V_{k_1} V_{k_2} V_{k_3} \int \td t_2  \td t_3  \mathcal{P}_F\nonumber \\
& \times\exp\left[ -\frac{i}{2}\left[k_1k_2\delta D_0^K(t_1-t_2)+k_2k_3\delta D_0^K(t_2-t_3)+k_3k_1\delta D_0^K(t_3-t_1)\right]\right] \nonumber \\
& \times\left. k_1^2  \cos\left[R_{21}+R_{31}\right] \sin\left[R_{12}+R_{32}\right]\sin\left[R_{23}+R_{13}\right] \right| _{t_1\to t}
\end{align}
with $s\equiv t_1-t'$, we have
\begin{align}
\Sigma^R(t)&=-4\sum_{k_1 k_2 k_3} V_{k_1} V_{k_2} V_{k_3} \int_{-\infty}^{\infty} \td s \exp\left[\frac{iF}{\eta}(k_1s+k_2t )\right]\nonumber \\
& \times\exp\left[ -\frac{i}{2}\left[k_1k_2\delta D_0^K(s-t)+k_2k_3\delta D_0^K(t)+k_3k_1\delta D_0^K(s)\right]\right] \nonumber \\
& \times  k_2k_3\cos\left[\frac{k_1k_2}{2}D_0^R(s-t)\right] \sin\left[ \frac{k_2k_3}{2}D_0^R(t)+\frac{k_3k_1}{2}D_0^R(s) \right]\sin\left[ \frac{k_1k_2}{2}D_0^R(t-s)+\frac{k_3k_1}{2}D_0^R(-s) \right]\nonumber \\
& -B\delta(t).
\end{align}
The mobility is 
\begin{align}
\mu(\omega=0)&=\frac{1}{i\eta^2} \int \td t  it\Sigma^R(t) \nonumber \\
&=-\frac{4}{\eta^2}\sum_{k_1 k_2 k_3} V_{k_1} V_{k_2} V_{k_3} \int_0^{\infty} t \td t \int_{-\infty}^{\infty} \td s \exp\left[\frac{iF}{\eta}(k_1s+k_2t )\right]\nonumber \\
&\qquad \qquad \times\exp\left[ -\frac{i}{2}\left[k_1k_2\delta D_0^K(s-t)+k_2k_3\delta D_0^K(t)+k_3k_1\delta D_0^K(s)\right]\right] \nonumber \\
&\qquad \qquad \times  k_2k_3\cos\left[\frac{k_1k_2}{2}D_0^R(s-t)\right] \sin\left[ \frac{k_2k_3}{2}D_0^R(t)+\frac{k_3k_1}{2}D_0^R(s) \right]\sin\left[ \frac{k_1k_2}{2}D_0^R(t-s)+\frac{k_3k_1}{2}D_0^R(-s) \right] \nonumber \\
&=-\frac{4}{\eta^2}\sum_{k_1 k_2 k_3} V_{k_1} V_{k_2} V_{k_3} \int_0^{\infty} t \td t \int_{0}^{\infty} \td s \exp\left[\frac{iF}{\eta}(k_1s+k_2t )\right]\nonumber \\
&\qquad \qquad \times\exp\left[ -\frac{i}{2}\left[k_1k_2\delta D_0^K(s-t)+k_2k_3\delta D_0^K(t)+k_3k_1\delta D_0^K(s)\right]\right] \nonumber \\
&\qquad \qquad \times  k_2k_3 \sin\left[ \frac{k_2k_3}{2}D_0^R(t)+\frac{k_3k_1}{2}D_0^R(s) \right]\sin\left[ \frac{k_1k_2}{2}D_0^R(t-s) \right] \nonumber \\
&-\frac{4}{\eta^2}\sum_{k_1 k_2 k_3} V_{k_1} V_{k_2} V_{k_3} \int_0^{\infty} t \td t \int_{0}^{\infty} \td s \exp\left[\frac{iF}{\eta}(-k_1s+k_2t )\right]\nonumber \\
&\qquad \qquad \times\exp\left[ -\frac{i}{2}\left[k_1k_2\delta D_0^K(s+t)+k_2k_3\delta D_0^K(t)+k_3k_1\delta D_0^K(s)\right]\right] \nonumber \\
&\qquad \qquad \times  k_2k_3\sin\left[ \frac{k_2k_3}{2}D_0^R(t) \right]\sin\left[ \frac{k_1k_2}{2}D_0^R(t+s)+\frac{k_3k_1}{2}D_0^R(s) \right] .
\end{align}
In the first term, the integration is restricted to $t>s$ due to the factor $D_0(t-s)$. Since $\int_0^{\infty} \td t\int_0^{\infty} \td s f(t,s) \Theta(t-s)=\int_0^{\infty} \td t\int_0^{\infty} \td s f(t+s,s) $, the first term is
\begin{align}
&-\frac{4}{\eta^2}\sum_{k_1 k_2 k_3} V_{k_1} V_{k_2} V_{k_3} \int_0^{\infty} (t+s) \td t \int_{0}^{\infty} \td s \exp\left[\frac{iF}{\eta}(k_2t-k_3s )\right]\nonumber \\
&\qquad \qquad \times\exp\left[ -\frac{i}{2}\left[k_1k_2\delta D_0^K(t)+k_2k_3\delta D_0^K(t+s)+k_3k_1\delta D_0^K(s)\right]\right] \nonumber \\
&\qquad \qquad \times  k_2k_3 \sin\left[ \frac{k_2k_3}{2}D_0^R(t+s)+\frac{k_3k_1}{2}D_0^R(s) \right]\sin\left[ \frac{k_1k_2}{2}D_0^R(t) \right].
\end{align}
As $k_2k_3t+k_1k_2(t+s)=-k_2(k_2t-k_1s)$, the total differential mobility reads
\begin{align}
\mu(\omega=0)&=\frac{4}{\eta^2}\sum_{k_1 k_2 k_3} V_{k_1} V_{k_2} V_{k_3} \int_0^{\infty} \td t \int_{0}^{\infty} \td s  k_2(k_2t-k_1s)\exp\left[\frac{iF}{\eta}(k_2t-k_1s )\right]\nonumber \\
&\qquad \qquad \times\exp\left[ -\frac{i}{2}\left[k_1k_2\delta D_0^K(t+s)+k_2k_3\delta D_0^K(t)+k_3k_1\delta D_0^K(s)\right]\right] \nonumber \\
&\qquad \qquad \times  \sin\left[ \frac{k_3k_1}{2}D_0^R(s)+\frac{k_1k_2}{2}D_0^R(t+s) \right]\sin\left[ \frac{k_2k_3}{2}D_0^R(t) \right]\nonumber \\
&=\frac{4}{\eta^2}\sum_{k_1 k_2 k_3} \int_0^{\infty} \td t \int_{0}^{\infty} \td s \left[ \Re[V_{k_1} V_{k_2} V_{k_3}] \cos\left[\frac{F}{\eta}(k_2t-k_1s )\right]-\Im[V_{k_1} V_{k_2} V_{k_3}] \sin\left[\frac{F}{\eta}(k_2t-k_1s )\right]\right]\nonumber \\
&\qquad \qquad \times k_2(k_2t-k_1s)\exp\left[ \frac{1}{\pi\eta}\left[k_1k_2 Q_2(t+s)+k_2k_3 Q_2(t)+k_3k_1Q_2(s)\right]\right] \nonumber \\
&\qquad \qquad \times  \sin\left[ \frac{k_3k_1}{2}D_0^R(s)+\frac{k_1k_2}{2}D_0^R(t+s) \right]\sin\left[ \frac{k_2k_3}{2}D_0^R(t) \right]
\end{align}
Thus, after relabeling the variables, the velocity is 
\begin{align}
v &=\int_0^F \td F \ \mu \nonumber \\
&= \frac{4}{\eta}\sum_{k_1 k_2 k_3} \int_0^{\infty} \td t_1 \int_{0}^{\infty} \td t_2 \left[ \Re[V_{k_1} V_{k_2} V_{k_3}] \cos\left[\frac{F}{\eta}(k_1t_1-k_3t_2 )\right]+\Im[V_{k_1} V_{k_2} V_{k_3}] \sin\left[\frac{F}{\eta}(k_1t_1-k_3t_2 )\right]\right]\nonumber \\
&\qquad \qquad \times k_1\exp\left[ \frac{1}{\pi\eta}\left[k_3k_1Q_2(t_1+t_2)+k_1k_2Q_2(t_1)+k_2k_3 Q_2(t_2)\right]\right] \nonumber \\
&\qquad \qquad \times  \sin\left[ \frac{k_1k_2}{2}D_0^R(t_1) \right]\sin\left[ \frac{k_2k_3}{2}D_0^R(t_2)+\frac{k_3k_1}{2}D_0^R(t_1+t_2) \right].
\end{align}
This is exactly same as eq.(\ref{Vinokurv3}) and eq.(8) in the main text.

\section{S5. Cancelation of $k$ summation in $v^{(3)}$ at low temperature}
As mentioned in the main text, the leading order contribution to the $v^{(3)}$ in the low temperature limit cancels out due to the summation $\sum_{k_1,k_2,k_3}$, which is shown in this section. Due to the momentum conservation $k_1+k2+k_3=0$ and invariance of the integrand under $k\to -k$, The leading order contributions come from three configurations; $(k_1,k_2,k_3) \times a/2\pi= (1,1,-2),(-2,1,1),(1,-2,1)$  

For $(k_1,k_2,k_3) \times a/2\pi= (1,1,-2)$, the integrand for the even function part with respect to $F$ reads
\begin{align}
&\   k_1  \left( \cos\left[ \frac{F}{\eta} \left(  k_1 t_1 - k_3 t_2\right) \right] -1\right) 
\sin \left[ \frac{1}{\pi\eta}  k_1 k_2 Q_1(t_1)\right]  
\sin \left[ \frac{1}{\pi\eta} \left(  k_2k_3 Q_1(t_2) + k_3 k_1Q_1(t_1+t_2) \right)  \right] \nonumber \\
& \quad \times    \exp\left[ \frac{1}{\pi\eta} \left(  k_1 k_2 Q_2(t_1) + k_2 k_3 Q_2(t_2) +k_3k_1Q_2(t_1+t_2)  \right) \right] \nonumber\\
&\sim   \left( \cos\left[ \frac{2\pi F}{\eta a} \left(  t_1 +2 t_2\right) \right] -1\right) 
\sin \left[ \frac{2}{\alpha}  Q_1(t_1)\right]  
\sin \left[ \frac{2}{\alpha} \left(  -2 Q_1(t_2) -2Q_1(t_1+t_2) \right)  \right] \nonumber \\
& \quad \times    \exp\left[ \frac{2}{\alpha} \left(  Q_2(t_1) -2 Q_2(t_2) -2Q_2(t_1+t_2)  \right) \right].
\end{align}
As the factor $(\cos[Ft]-1)$ gives a small $t$ cutoff and due to the exponential factors, the leading contribution comes from the region $t_1\sim T_F \equiv \frac{\eta a}{2\pi F}$ and $t_2=t\ll t_1$. In this region, since $Q_1(t_1+t_2)\sim Q_1(t_1)\sim\pi/2$ and $Q_2(t_1+t_2)\sim Q_2(t_1) =Q_2(T_F)$, we have
\begin{align}
&\sim -
\sin \left[ \frac{\pi}{\alpha} \right]  
\sin \left[ \frac{2}{\alpha} \left(  2 Q_1(t) +\pi  \right)  \right]  \exp\left[ \frac{2}{\alpha} \left( -2 Q_2(t) -Q_2(T_F)  \right) \right].
\end{align}

For $(k_1,k_2,k_3) \times a/2\pi= (-2,1,1)$, similarly we have
\begin{align}
&\sim  -2 \left( \cos\left[ \frac{2\pi F}{\eta a} \left( 2 t_1 + t_2\right) \right] -1\right) 
\sin \left[ -\frac{4}{\alpha}  Q_1(t_1)\right]  
\sin \left[ \frac{2}{\alpha} \left(   Q_1(t_2) -2Q_1(t_1+t_2) \right)  \right] \nonumber \\
& \quad \times    \exp\left[ \frac{2}{\alpha} \left(  -2Q_2(t_1) +  Q_2(t_2) -2Q_2(t_1+t_2)  \right) \right] \nonumber \\
&\sim  -2 
\sin \left[ \frac{4}{\alpha}  Q_1(t)\right]  
\sin \left[ \frac{\pi}{\alpha}   \right] 
\exp\left[ \frac{2}{\alpha} \left(  -2Q_2(t)  -Q_2(T_F)  \right) \right] 
\end{align}
by setting  $t_2\sim T_F $ and $t_1=t\ll t_2$.

For $(k_1,k_2,k_3) \times a/2\pi= (1,-2,1)$, there are two regions which give the leading order contributions corresponding to two blue regions in Fig.2 in the main text. One is $t_1\sim T_F $ and $t_2=t\ll t_1$, where
\begin{align}
&\sim  \left( \cos\left[ \frac{2\pi F}{\eta a} \left(  t_1 - t_2\right) \right] -1\right) 
\sin \left[ -\frac{4}{\alpha}  Q_1(t_1)\right]  
\sin \left[ \frac{2}{\alpha} \left(  -2 Q_1(t_2) +Q_1(t_1+t_2) \right)  \right] \nonumber \\
& \quad \times    \exp\left[ \frac{2}{\alpha} \left(  -2Q_2(t_1) -2  Q_2(t_2) +Q_2(t_1+t_2)  \right) \right] \nonumber \\
&\sim   
\sin \left[ \frac{2\pi}{\alpha} \right]  
\sin \left[ \frac{2}{\alpha} (2Q_1(t_2) -\pi/2)   \right] 
\exp\left[ \frac{2}{\alpha} \left(   -2  Q_2(t) -Q_2(T_F)  \right) \right] .
\end{align}

The other contribution comes from the region $t_2\sim T_F $ and $t_1=t\ll t_2$, where
\begin{align}
\sim 
\sin \left[ \frac{4}{\alpha}  Q_1(t)\right]  
\sin \left[ \frac{\pi}{\alpha}  \right] \exp\left[ \frac{2}{\alpha} \left(  -2Q_2(t) -  Q_2(T_F)  \right) \right] .
\end{align}

Summing up all the contributions, we see the cancellation as
\begin{align}
& \int _0^\infty \td t
\left( -
\sin \left[ \frac{\pi}{\alpha} \right]  
\sin \left[ \frac{2}{\alpha} \left(  2 Q_1(t) +\pi  \right)    \right]
-
\sin \left[ \frac{\pi}{\alpha}   \right] 
\sin \left[ \frac{4}{\alpha}  Q_1(t)\right]    
+
\sin \left[ \frac{2\pi}{\alpha} \right]  
\sin \left[ \frac{2}{\alpha} (2Q_1(t) -\pi/2)   \right]  \right)    \nonumber \\
&\qquad \qquad  \times  \exp\left[ \frac{2}{\alpha} \left(  -2Q_2(t) -  Q_2(T_F)  \right) \right] \nonumber \\
&= \int _0^\infty \td t  \sin \left[ \frac{4}{\alpha}  Q_1(t)\right]   
\left( -
\sin \left[ \frac{\pi}{\alpha} \right]  
\cos \left[ \frac{4\pi}{\alpha}  \right]
-
\sin \left[ \frac{\pi}{\alpha}   \right]  
+
\sin \left[ \frac{2\pi}{\alpha} \right]  
\cos \left[ \frac{\pi}{\alpha}  \right]  \right)    \exp\left[ \frac{2}{\alpha} \left(  -2Q_2(t) -  Q_2(T_F)  \right) \right]\nonumber \\
&=0	
\end{align}
where we have used a identity $\int_0^\infty  \td t \cos \left[ \frac{4}{\alpha} Q_1(t)\right]\exp\left[ -\frac{4}{\alpha} Q_2(t)\right] =0$ \cite{LeggettRMP}.

\begin{figure}[!t]
	\includegraphics[width=0.5\textwidth]{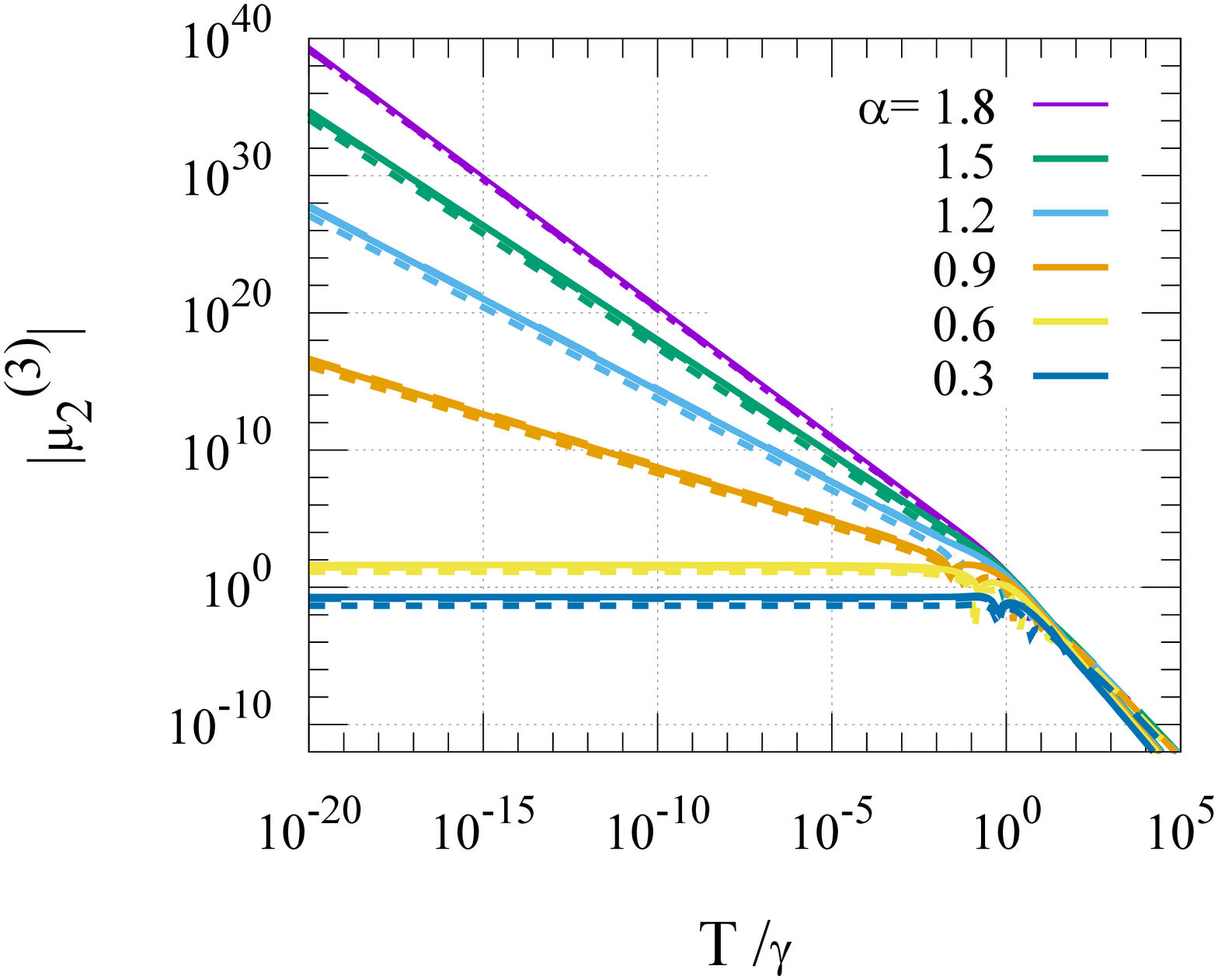}
	\caption{ 
		$\mu_2^{(3)}$ value before the $k_1,k_2,k_3$ summation vs $T$ for all the temperature regions down to ultra low temperature. Three curves (line, broken line dotted line) are for $(k_1,k_2,k_3)\propto(-2,1,1),(1,1,-2),(1,-2,1)$, respectively.
	}
	\label{Fig:cancel}
\end{figure}

The cancellation is numerically checked. In Fig.\ref{Fig:cancel} we shows the $\mu_2^{(3)}$ values before the $k_1,k_2,k_3$ summation down to ultra low temperature. Three curves for each value of $\alpha$ corresponds to $(k_1,k_2,k_3)\propto(-2,1,1),(1,1,-2),(1,-2,1)$ terms. All the lines follow the asymptotic behavior $\sim T^{2/\alpha-3}$. We have confirmed the cancellation occurs after summing up these three terms up to $12$ digits which is close to the limitation of the double precision calculation.

On the other hand in the {\it low temperature region} discussed in the main text $\mu_2^{(3)} \propto T^{6/\alpha-4}$, cancellation is incomplete and numerical results are reliable.

\section{S6. High temperature limit of $\mu_2^{(3)}$}
In this section, we derive the high temperature power law decay; $\mu_2^{(3)}\sim T^{-11/4}$ according to the analysis in ref. \cite{Vinokur} where $\mu_2^{(3)}$ is the second order mobility calculated based on the expansion of the eq.(8) in the main text with respect to $F$. In high temperature limit, the small values of $t_1,t_2$ give a dominant contribution to the integral since the integrand decays immediately due to the factor $e^{-Tt}$. In this limit we can expand $Q_1$ and $Q_2$ as
\begin{align}
Q_1(t) &= \gamma t -\frac{1}{3} \gamma^3 t^3 +o(t^5)\\
Q_2(t) &= \frac{T}{\gamma} \left( \gamma^2 t^2 -\frac{1}{6}\gamma^4t^4 \right) + o(t^6)
\end{align}
Introducing new variable as
\begin{equation}
t_\pm = k_1t_1\pm k_3 t_2,
\end{equation}
the integral range changes according to the relative sign between $k_1$ and $k_3$;
\begin{align}
\mathrm{(i)} \ \ & t_+ >0, \quad -t_+ < t_-<t_+ \quad (\mathrm{for}\ k_1k_3>0)\\
\mathrm{(ii)}\ \ & t_- >0, \quad -t_- < t_+<t_- \quad (\mathrm{for}\ k_1k_3<0).
\end{align}

The exponential factor in the integrand is expanded as
\begin{align}
E&\equiv \frac{2}{\alpha} \left[ k_1k_2Q_2(t_1)+k_2k_2Q_2(t_2)+k_3k_1Q_2(t_1+t_2) \right] \nonumber \\
&=-T\left[ at_-^2 + b t_-t_+^3 +c t_+^4 \right] + o(t_-^3  ,   t_-^2t_+   ,   t_-t_+^4  ,  t_+^5)
\end{align}
and other factor is
\begin{align}
F&\equiv (k_1t_1-k_3t_3)^2 \sin \left[ \frac{1}{\pi\eta}  k_1 k_2 Q_1(t_1)\right]  \sin \left[ \frac{1}{\pi\eta} \left(  k_2 k_3Q_1(t_2) + k_3k_1 Q_1(t_1+t_2) \right)  \right]  \nonumber \\
&= dt_-^4 +et_-^3 t_+ + ft_-^2t_+^4 +o(t_-^5  ,  t_-^4t_+  ,  t_-^3t_+^2  ,  t_-^2 t_+^5)
\end{align}
with appropriate real coefficients $a,b,c,d,e,f$. As the factor $F $ is some power of $t_+$ and $t_-$, by introducing the transformation of the variable $y\equiv T^{1/2}t_-$, temperature dependence can be factored out. Appropriate transformation of $t_2$ depends on the situations (i) and (ii). 

For the case (i), the integral region of $t_+$ is bounded by $\pm t_-$ therefore the force order terms in the expansion of $E$ can be neglected, the integral reads
\begin{align}
\mu_2^{(3)} &\sim \int \td t_1 \int \td t_2 \ F e^{-E}\nonumber \\
&\sim \int_0^\infty  \td t_- \int_{-t_-}^{t_-} \td t_+ e^{-aTt_-^2 } (dt_-^4 +et_-^3 t_+ + ft_-^2t_+^4) \nonumber \\
&\sim \int_0^\infty  \td t_-  e^{-aTt_-^2 } (d't_-^5  + ft_-^7) \nonumber \\
&= \int_0^\infty  \frac{\td y}{T^{1/2}}  e^{-ay^2 } \left(d' \left[\frac{ y}{T^{1/2}}\right]^5 + f\left[\frac{ y}{T^{1/2}}\right]^7\right) \sim T^{-3} .
\end{align}
This power law is dominated by the case (ii). In this case the integral over $t_+$ diverges unless we include the forth order term in the expansion of $E$. In this case, if we set $z\equiv T^{1/4} t_+$, the second term in $E\sim ay^2 + bT^{-1/4}z^3+cz^4$ vanishes in high temperature limit. Thus we have
\begin{align}
\mu_2^{(3)} &\sim \int_0^\infty  \td t_+ \int_{-t_+}^{t_+} \td t_-  e^{-aTt_-^2 -cTt_+^4} (dt_-^4 +et_-^3 t_+ + ft_-^2t_+^4) \nonumber \\
&\sim \int_0^\infty  \frac{\td z}{T^{1/4}} \int_{-zT^{1/4}}^{zT^{1/4}} \frac{\td y}{T^{1/2}}  e^{-ay^2 -cz^4} \left(d\left[\frac{y}{T^{1/2}}\right]^4 +e\left[\frac{y}{T^{1/2}}\right]^3 \left[\frac{z}{T^{1/4}}\right] + f\left[\frac{y}{T^{1/2}}\right]^2\left[\frac{z}{T^{1/4}}\right]^4 \right) \sim T^{-11/4}
\end{align}
which dominates in the high temperature limit of $\mu_2^{(3)}$.
The discrepancy between the Scheidl-Vinokur's result $\mu_2^{(3)}\sim T^{-17/6}$ and ours $T^{-11/4}$ is attributed to the choice of the cutoff function. The former result is obtained by the Lorentz cutoff while we use the exponential cutoff for the evaluation of $Q_1$ and $Q_2$.

In completely parallel discussion, for the general even order mobility, we see $\mu_{2n}^{(3)} \sim T^{-7/4-n}$. This is clear from that the expansion of $F$ include the prefactor $t_-^{2n}$ for $\mu_{2n}^{(3)} $.

\end{document}